\documentclass{article}
\usepackage[utf8]{inputenc}
\usepackage{authblk}
\usepackage[sort&compress]{natbib}
\usepackage{graphicx}
\usepackage{tabularx}
\usepackage{amsmath}
\usepackage{upgreek}
\usepackage{hyperref}
\hypersetup{colorlinks, citecolor=blue, filecolor=black, linkcolor=black, urlcolor=black}
\usepackage{mathtools}
\usepackage{amssymb}
\usepackage{nomencl}
\usepackage{geometry}
\usepackage{epstopdf}
\usepackage{subcaption}
\usepackage{xcolor}
\usepackage{cleveref}
\usepackage[export]{adjustbox}
\usepackage{wrapfig}
\usepackage{float}
\usepackage{nameref}
\usepackage[ruled,vlined]{algorithm2e}
\usepackage{soul}
\DeclareGraphicsExtensions{.png,.pdf}
\usepackage[format=plain,font=it]{caption}
\usepackage[multiple]{footmisc}
\usepackage{bigfoot}
\usepackage{multirow}
\usepackage{lineno}

\title{Variational PINNs with tree-based integration and boundary element data in the  modeling of multi-phase architected materials}

\author[1]{Dimitrios C. Rodopoulos\thanks{Corresponding Author: Dimitrios C. Rodopoulos, E-mail:  \textcolor{blue}{ d.rodopoulos@nyu.edu}}}
\author[1]{Panos Pantidis}
\author[1,2]{Nikolaos Karathanasopoulos}
\affil[1]{{\small New York University AD, Department of Mechanical Engineering,
 Abu Dhabi, UAE}}
\affil[2]{{\small New York University, Department of Mechanical and Aerospace Engineering, Tandon School of Engineering, Brooklyn, NY, 11201, USA}}

\DeclareMathOperator*{\argminA}{arg\,min}

\newcommand{\Div}{\nabla \cdot}

\newcommand{\nv}{\mathbf{n}}

\date{}
\providecommand{\keywords}[1]
{
	\small	
	\textbf{\textit{Keywords:}} #1
}
\begin{document}

\maketitle
\vspace{-3em}
{
\section*{Abstract}
The current contribution develops a Variational Physics-Informed Neural Network (VPINN)-based framework for the analysis and design of multiphase architected solids. The elaborated VPINN methodology is based on the Petrov-Galerkin approach, with a deep neural network acting as trial function and local polynomials as test functions. For the analysis, a Galerkin Boundary Element Method (GBEM) scheme is developed to generate the mechanical field data, employing solely domain boundary information. The VPINN methodology is complemented by an adaptive, tree-based integration scheme for the evaluation of the weak-form integrals. Different double-phase material architectures are considered, with the VPINNs demonstrating their ability to capture the deformation fields with considerable accuracy. Moreover, the performance enhancement by the incorporation of additional semi-analytical information at auxiliary internal points is analyzed. Tree-based integration schemes are shown to be capable of robustly capturing inner material discontinuities upon substantial computational cost reductions. The results suggest that the proposed VPINN formulation offers comparative advantages in the modeling of multiphase architected materials compared to classical PINN formulations. The analysis paves the way for the development of variational physics-informed computational models for the mechanical analysis of complex architected multiphase materials and structures.
\\
\\
\keywords{Physics-Informed Neural Networks, Variational formulation, Architected materials, Boundary Element method, Machine Learning modeling}
}


\section{Introduction}
{
\hspace{\parindent}
Recently, Deep Learning (DL) methods have gained a lot of attention in the modeling of diverse physical phenomena and processes \cite{RAISSI2019686}. Artificial Neural Networks (ANNs), a specific class of DL methods, have emerged as a prominent tool in the solution of differential equations for a wide range of engineering applications \cite{RAISSI2018125, 712178, BERG201828, HU2024112495,LECLEZIO2024102260,https://doi.org/10.1002/nme.7176}. Physics-informed neural networks (PINNs) constitute a deep learning framework that is based on ANNs, where the physical phenomena are formulated as an optimization task that incorporates information coming from the underlying partial differential equations (PDEs) in the model loss function. 

One of the key aspects for the design of robust PINN models is the development of appropriate loss functions that allow for a robust optimization of the quantities of interest \cite{RAISSI2019686}. To this end, several physics-based deep learning frameworks have been up to now developed, including Deep Galerkin Method (DGM) models based on least squares \cite{SIRIGNANO20181339}, Physics-Informed Neural Networks based on collocation methods \cite{RAISSI2019686}, as well as Deep Energy Methods (DEM) that exploit energy functionals \cite{SAMANIEGO2020112790,NGUYENTHANH2020103874,NGUYENTHANH2021114096,Chadha2023}. Variational physics-informed neural networks have typically been based on Galerkin formulations, thus on weighted-residual approximations of the PDEs \cite{DBLP:journals/corr/abs-1912-00873,pmlr-v120-khodayi-mehr20a,KHARAZMI2021113547,sikora2024comparison,rojas2024} . Various engineering applications, such as fluid mechanics \cite{RAISSI2019686,doi:10.1137/18M1191944}, electromagnetics \cite{SON2023102035,ZHANG2025113798}, and solid mechanics \cite{HAGHIGHAT2021113741,JEONG2023115484,MANAV2024117104,FUHG2022110839,li2024finitepinnphysicsinformedneuralnetwork,BASTEK2023104849,https://doi.org/10.1002/nme.6828} use PINNs as surrogate models \cite{RAISSI2019686,HAGHIGHAT2021113741,MANAV2024117104,li2024finitepinnphysicsinformedneuralnetwork,https://doi.org/10.1002/nme.6828}, inversion schemes \cite{RAISSI2019686,HAGHIGHAT2021113741,li2024finitepinnphysicsinformedneuralnetwork}, or as a basis for the acceleration of finite element computations \cite{PANTIDIS2023115766, PANTIDIS2024116940, PANTIDIS2025INTEGRATED}. 

Among the various engineering applications of PINNs, solid mechanics constitutes a key area of importance, due to its direct relevance to a wide range of material and structural engineering problems. Multiphase composites are used in diverse engineering fields for the design of structural components in aerospace, automotive, and biomedical engineering applications \cite{SALEH2020108376,AMReview,El-Galy2019,C8MH00653A}. Advanced materials typically employ complex internal architectures, requiring accurate and robust numerical methods for their analysis. The Finite Element Method (FEM) has been the primary tool for the numerical modeling of multiphase materials \cite{DALAQ2016169,ZHANG2021113693,LI2020108229}, noting its efficiency and flexibility in the treatment of complex internal boundaries and material nonlinearities \cite{RODOPOULOS2019160, RODOPOULOS2020259}. Complementary to volume discretization techniques \cite{RODOPOULOS2020259}, boundary discretization methods, namely Boundary Element Methods (BEM) have been, as-well, developed \cite{CHEN2005513,GORTSAS2018103,WU2021245}. The latter offers out of their formulation significant advantages in the reduction of the computational cost for topologies with high surface to volume ratios \cite{RODOPOULOS_2021,RODOPOULOS2021118,doi:10.1142/S1756973717400017}. High-fidelity methods, such as FEM and BEM models, can be used to generate target (labeled) data for the solution of the underlying PDEs than can be used for the training of the PINN models \cite{LECLEZIO2024102260,HAGHIGHAT2021113741,TeoPINNs}.

Different PINN formulations have been developed for interface problems in heterogeneous materials \cite{doi:10.1126/sciadv.abk0644,https://doi.org/10.1002/nme.7388,DIAO2023116120,SARMA2024117135,LINGHU2025116223}. More specifically, PINNs for the inverse identification of the internal structure of materials with a secondary inner architecture or a defect have been developed \cite{doi:10.1126/sciadv.abk0644}, as well as mixed formulations for the thermoelastic analysis of heterogeneous domains \cite{https://doi.org/10.1002/nme.7388}. Furthermore, mixed formulations have found application in the modeling of layered domains with an inner material distribution \cite{DIAO2023116120}, employing domain decomposition schemes. Dedicated PINN models for the solution of Poisson's type equations in domains with inner heterogeneous material interfaces have been elaborated \cite{SARMA2024117135}, demonstrating substantial flexibility in the modeling of inner material discontinuities. Moreover, multi-scale, higher-order PINN-based formulations for the computation of the deformed displacement field of composites with secondary inner circular phases have been elaborated \cite{LINGHU2025116223}.

However, despite the proven ability of variational techniques to treat rough input data that include singularities and sharp changes \cite{DBLP:journals/corr/abs-1912-00873, KHARAZMI2021113547}, they have not been up to now explored in solid mechanics applications with multiphase architected materials. It should be noted that the increased computational demands appearing in such problems, require efficient ways for the accurate integration of the physics-based terms in the model loss function, beyond Riemann sums \cite{NGUYENTHANH2020103874,https://doi.org/10.1002/nme.7388}, Trapezoidal and Simpson \cite{NGUYENTHANH2020103874}, or Gauss-Legendre rules \cite{NGUYENTHANH2021114096}. Gauss-Legendre rules favor accurate numerical integration, but require simple domains, such as rectangular grids or patterns in which simple coordinate transformations can be applied. Nevertheless, for complex domains appearing in architected materials, accurate numerical integration remains a challenging task, requiring a large number of collocation points for the reduction of integration errors in uniformly distributed meshes, such as in Delaunay triangulation meshes \cite{FUHG2022110839}. As such, techniques that provide sufficiently accurate geometric approximations, while minimizing the associated computational cost, are required.

The present study elaborates a Variational PINN formulation for the computation of the mechanical response of architected materials with inner phase discontinuities. The method is based on a local Petrov-Galerkin scheme, where instead of the differential operator, the weak formulation of the boundary value problem is enforced to be satisfied in the interior of the modeling domain, using local piece-wise linear polynomials as test functions. For the data generation, a Galerkin Boundary Element (GBEM) scheme that can semi-analytically compute the mechanical response of domain boundaries, interfaces, and different internal points, is employed \cite{RODOPOULOS2025113159}. For the accurate numerical evaluation of the weak form integrals, a tree-based adaptive integration scheme is developed, which allows for the partition of the domain into a set of adaptively-sized integration cells using a quadtree (2D) clustering approach, in which the Gauss-Legendre integration rule is applied. Noting that the data generation and the domain subdivision processes are based solely on the boundary and interface geometry, the meshless nature of the VPINN formulation is preserved, providing the flexibility to work on multi-phase materials with complex inner architectures.

The paper is organized as follows: The methodology is elaborated in Section \ref{Sec:methods}, providing a thorough description of the strong and weak formulations of the boundary value problem (Section \ref{Sec:ProblemStatement}), along with the VPINN formulation (Section \ref{Sec:VarPINN}) and the overarching machine learning framework (Section \ref{Sec:ML_Frame}). Section \ref{Sec:Results} illustrates the numerical results on the different test cases and the evaluation of the sensitivity and computational efficiency. Finally, a discussion on the conclusions and future outlook is presented in Section \ref{Sec:Conclusions}.

}

\begin{table}[h]
	\begin{center}
		\begin{tabular}{ l p{7cm} l p{5cm} l p{5cm} l }
        \hline
               \multicolumn{4}{c}{\textbf{Nomenclature}} 
                \\
               \multicolumn{2}{l}{\textbf{List of symbols}}
               & \multicolumn{2}{l}{~}
                \\                
			$\mathbb{R}^{d}$ & d-dimensional Euclidean space  &$\mathcal{N}(\mathbf{x};\boldsymbol{\lambda})$  & Neural Network with parameters $\lambda$
                \\                 
			$\mathbf{x}$ & Position vector in $\mathbb{R}^{d}$  &$H_{\lambda}$  & Space of parameters $\lambda$          
                \\                                 
			$\hat{\mathbf{x}}_{i}$ & Cartesian unit basis vector &$\boldsymbol{\lambda}$  & Trainable parameters            
                \\ 
			$\cdot$ & Dot product  &$\mathbf{W}$  & Trainable weight matrix
                \\ 
			$\times$ & Cross product  &$\mathbf{b}$  & Trainable bias vector
                \\                 
			$\otimes$ & Tensor product  &$g^{i}$  & Activation function of the $i$th layer
                \\                                 
			$:$ & Double dot product  &$P_{\Omega}$  & Set of differential operator collocation points
			\\
			$\Tilde{\mathbf{I}}$ & Unit tensor  &$D_{\partial\Omega}$  & Set of boundary training data
			\\ 
			$\nabla$ & Gradient operator  &$D_{\Omega}$  & Set of domain training data
			\\            
			$\nabla^{2}$ & Laplacian Operator  &$E$  & Integration cell
			\\
			$\partial$ & Boundary operator  &$\mathcal{T}(\Omega_{i})$  & Set of integration cells of a domain $\Omega_{i}$
			\\            
			$\partial_{n}$ & Directional derivative on the direction of $\mathbf{n}$   &$(\boldsymbol{\xi},h(\boldsymbol{\xi}))$  & Pair of integration point and the corresponding weight
			\\   
			$\mathbf{n}$ & Unit normal vector  &$\mathcal{G}(E)$  & Set of integration points/weights $(\boldsymbol{\xi},h(\boldsymbol{\xi}))$ located in $E$
			\\  
			$\mathbf{u}$ & Displacement vector &$\mathcal{G}(\Omega_{0})$  & Set of integration points/weights $\mathcal{T}(\Omega_{0}) = \bigcup E$
                 \\
			$\boldsymbol{\Sigma}$ & Stress tensor &$\mathcal{Q}(\Omega_{0})$  & Quadtree of $\Omega_{0}$
                 \\
			$\mathbf{E}$ & Strain tensor &$S(E)$  & Sons of a parent cell $E$, $S(E) = \{\tau_{1},\tau_{2},\tau_{3},\tau_{4} \}$
            \\
			$\uplambda$, $\upmu$ & Lame's parameters &~  & ~             
                 \\
			$\Omega_{0}$ & Domain of the architected material &\multicolumn{2}{l}{\textbf{Abbreviations}}
                \\
			$\Omega_{i}$ & Domain of the $i$th material phase &PINN  & Physics-Informed Neural Network
                 \\
			$\Gamma_{I}$ & Interface boundary &VPINN  & Variational Physics-Informed Neural Network
                 \\
			$\sqcup$ & Disjoint set union &VFEM  & Voxel Finite Element Method
                 \\
			$\#D$ & Number of elements of a set $D$  &~  & ~
                 \\
			$L^{2}({\Omega})$ & Lebesque space: $\{ f:\Omega \rightarrow \mathbb{R} | \int_{\Omega}f^{2}d\Omega < \infty\}$  &~  &       
            \\
			$H^{1}({\Omega})$ & Sobolev space: $\{ f \in L^{2}({\Omega}) | \frac{\partial f}{\partial x_{j}} \in L^{2}({\Omega}) \text{, } j = 1,2,...,d \}$  &~  & ~            
            \\
			$\mathbf{w}$, $\boldsymbol{\upsilon}$ & Vector test functions  &~  & ~ 
            \\
			$\mathcal{W}_{\Omega}(\mathbf{u};\mathbf{w})$ & Weak formulation of the elastostatic problem  &~  & ~ 
            \\
			$\mathcal{U}_{\Omega}(\mathbf{x}')$ & Elastostatic integral operator  &~  & ~ 
            \\            
			$\tilde{\mathbf{U}}$, $\tilde{\mathbf{T}}$ & Integral kernels of the elastostatic problem  &~  & ~                               
            \\              
			\hline
		\end{tabular}
	\end{center}
	\label{tab:nomenclature}
\end{table}

\clearpage

\section{Methodology} 
\label{Sec:methods}
{

\subsection{Solid mechanics of multi-phase materials}
\label{Sec:ProblemStatement}

\hspace{\parindent}
An architected multi-phase material domain $\Omega_{0}$ formed by the subdomains $\Omega_{1}$ and $\Omega_{2}$ is considered, with base material properties $E^{1}$, $\nu^{1}$ (Young's modulus and Poisson's ratio, $\Omega_{1}$, Fig.\,\,\ref{fig:bvp}), and $E^{2}$, $\nu^{2}$ ($\Omega_{2}$, Fig.\,\,\ref{fig:bvp}). The second material phase is defined as the union of $N$ non-intersecting simply-connected domains $\Omega_{2}^{i}$ ($i=1,2,...,N$), i.e: $\Omega_{2} = \bigcup_{i=1} \Omega_{2}^{i}$, while the boundary of each domain is given as $\Gamma_{1} = \partial\Omega_{0} \cup \Gamma_{I}^{1}$, and $\Gamma_{2} = \Gamma_{I}^{2}$ (Fig.\ref{fig:bvp}). Accordingly, the interface boundary between the two material phases, $\Omega_{1}$ and $\Omega_{2}$ is defined as $\Gamma_{I} = \Gamma_{I}^{1} \cap \Gamma_{I}^{2}$. Considering a homogeneous and linearly isotropic material for each of the phases and in the absence of distributed elastic body loads, the elastostatic displacement field $\mathbf{u}:\mathbf{x} \in \Omega_{0}\longmapsto (u_{1}(\mathbf{x}),u_{2}(\mathbf{x}))$ satisfies the following system of equilibrium, constitutive and kinematic equations \cite{EslamiBook}:
\begin{equation} \label{eq:21e1}
\begin{aligned}
\begin{matrix}
\Div \boldsymbol{\Sigma}(\mathbf{u}) = \mathbf{0}
\\
\boldsymbol{\Sigma}(\mathbf{u}) = \uplambda \text{tr} \big[\mathbf{E}(\mathbf{u})\big] \Tilde{\mathbf{I}} + 2\upmu\mathbf{E}(\mathbf{u})
\\
\mathbf{E}(\mathbf{u}) = \frac{1}{2}\big(\nabla \mathbf{u} + \mathbf{u} \nabla)
\end{matrix}
, \hspace{0.2cm} \mathbf{x} \in \Omega_{0} \hspace{0.2cm}
\end{aligned}
\end{equation}
where:
\begin{equation} \label{eq:21e2}
\begin{aligned}
\uplambda &= \frac{\mathrm{E} \upnu}{(1+\upnu)(1-2\upnu)}
,  \hspace{0.1cm}
\upmu &= \frac{\mathrm{E}}{2(1+\upnu)}
\end{aligned}
\end{equation}
and $\boldsymbol{\Sigma}(\mathbf{u}) = \Sigma_{ij}(\mathbf{u})\hat{\mathbf{x}}_{i} \otimes \hat{\mathbf{x}}_{j}$ and $\mathbf{E}(\mathbf{u}) = E_{ij}(\mathbf{u})\hat{\mathbf{x}}_{i} \otimes \hat{\mathbf{x}}_{j}$ denote the stress and strain second-rank tensors, respectively, with $\otimes$ being the tensor product operator and $\hat{\mathbf{x}}_{i}$ $(i=1,2)$ representing the Cartesian unit vectors. In Eq. \eqref{eq:21e1}, $\nabla$ stands for the nabla operator, $\text{tr}(\cdot)$ for the trace operator, while $\Tilde{\mathbf{I}} = \hat{\mathbf{x}}_{1} \otimes \hat{\mathbf{x}}_{1} + \hat{\mathbf{x}}_{2} \otimes \hat{\mathbf{x}}_{2}$ for the unit tensor. The interface  boundary $\Gamma_{I}$ as well as the exterior boundary $\partial\Omega_{0} = \Gamma_{D} \cup \Gamma_{N}$ are subject to the following conditions:
\begin{equation}\label{eq:21e3}
\begin{aligned}
\left.\begin{matrix}
\mathbf{u}^{1} = \mathbf{u}^{2}
\\ 
\mathbf{t}^{1}(\mathbf{u}) + \mathbf{t}^{2}(\mathbf{u}) = \mathbf{0}
\end{matrix}\right\}
, \hspace{0.1cm} \mathbf{x} \in \Gamma_{I}
  \hspace{0.5cm}
\begin{matrix}
\mathbf{u} = \mathbf{u}^{0}, \hspace{0.1cm} \mathbf{x} \in \Gamma_{D}
\\ 
\mathbf{t}(\mathbf{u}) = \mathbf{t}^{0}, \hspace{0.1cm} \mathbf{x} \in \Gamma_{N}
\end{matrix},
\hspace{0.2cm}
\partial \Omega_{0} = \Gamma_{D} \cup \Gamma_{N}
\end{aligned}
\end{equation}
where $\Gamma_{D}$ and $\Gamma_{N}$ correspond to the boundary parts where Dirichlet and Neumann boundary conditions are prescribed, respectively. In Eq. \eqref{eq:21e3}, the vector $\mathbf{t}(\mathbf{u}) = \nv \cdot \boldsymbol{\Sigma}(\mathbf{u})$ stands for the surface traction, and $\cdot$ and $\times$ represent the dot and cross products, respectively. The weak formulation of the solid mechanics problem is obtained by multiplying with an arbitrary vector function $\mathbf{w}: \Omega_{0} \rightarrow \mathbb{R}^{2}$: 
\begin{equation}  \label{eq:21e4a}
\begin{aligned}
\int_{\Omega_{0}} \mathbf{w} \cdot \Big[ \Div \boldsymbol{\Sigma}(\mathbf{u}) \Big] d\Omega = 0
\end{aligned}
\end{equation}
Considering the vector identity $\mathbf{w} \cdot \big[ \Div \boldsymbol{\Sigma}^{T} \big] = \Div \big[\mathbf{w}\cdot \boldsymbol{\Sigma}\big] - \nabla \mathbf{w}:\boldsymbol{\Sigma}$ and the symmetry relation $\boldsymbol{\Sigma}^{T} = \boldsymbol{\Sigma}$, and implementing the divergence theorem, we obtain the following Petrov-Galerkin weak formulation for a pair of Hilbert spaces $V$ and $W$:
\begin{equation}  \label{eq:21e4b}
\begin{aligned}
&\text{Find} \hspace{0.1cm} \mathbf{u} \in V \hspace{0.1cm} \text{such that}
\hspace{0.1cm}\\
&\mathcal{W}_{\Omega_{0}}(\mathbf{u};\mathbf{w}) = \int_{\partial\Omega_{0}} \mathbf{w} \cdot \mathbf{t}(\mathbf{u}) d\Gamma - \int_{\Omega_{0}} \nabla \mathbf{w} : \boldsymbol{\Sigma}(\mathbf{u}) d\Omega = 0 \hspace{0.1cm},  \forall \hspace{0.1cm} \mathbf{w} \in W
\end{aligned}
\end{equation}
As the test function $\mathbf{w}$ should be chosen in such a way that the Dirichlet conditions on $\Gamma_{D}$ are satisfied, the space $W$ can be defined as the following Sobolev space $W = \{ \mathbf{w} \in [H^{1}(\Omega_{0})]^{2} \big | \mathbf{w} = \mathbf{0} \hspace{0.1cm} \text{on} \hspace{0.1cm} \Gamma_{D}\}$ \cite{evans10}. The definition of the solution space $V$ depends on the choice of the trial function. An appropriate space, that is the basis to define a finite element method is: $V = \{ \mathbf{u} \in [H^{1}(\Omega_{0})]^{2} \big | \mathbf{u} = \mathbf{u}_{0} \hspace{0.1cm} \text{on} \hspace{0.1cm} \Gamma_{D}\}$.
\begin{figure}[H]
\centering
\includegraphics[scale=0.7]{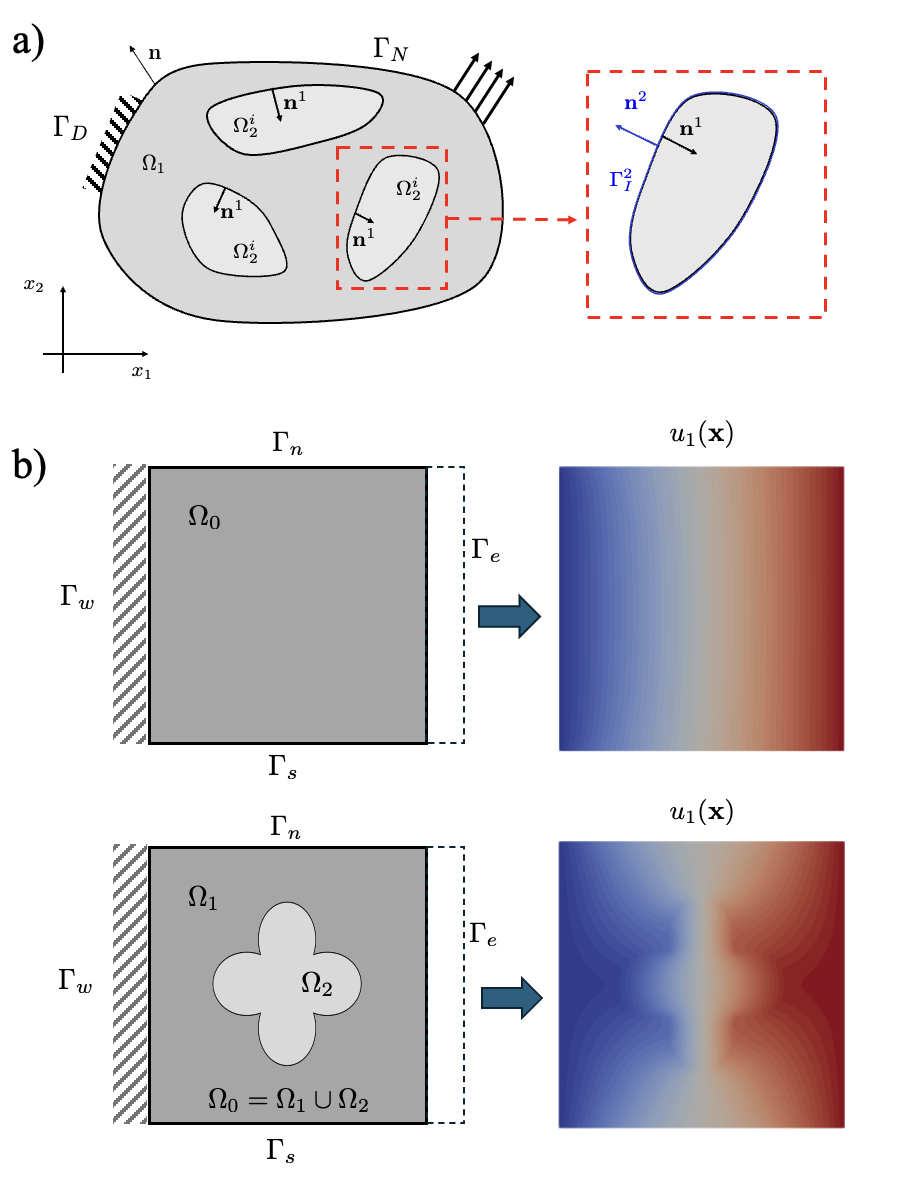} 
\caption{a) A double-phase material $\Omega_{0}$, complemented by the phases $\Omega_{1}$ and $\Omega_{2}$, and b) a square plate $\Omega_{0}$ without ($\Omega_{2} = \emptyset$) and with ($\Omega_{2} \neq \emptyset$) second phase material, subject to the Dirichlet (displacement) boundary conditions on $\Gamma_{D} = \Gamma_{w} \cup \Gamma_{e}$ as well as the Neumann (traction) boundary conditions on $\Gamma_{N} = \Gamma_{n} \cup \Gamma_{s}$. }
\label{fig:bvp}
\end{figure}

\subsection{Variational Physics Informed Neural Networks (VPINN)}
\label{Sec:VarPINN}

In this sequel, the machine learning based Variational Physics Informed Neural Network (VPINN) formulation is presented.
\subsubsection{Formulation}
\label{Sec:vpinn}
\hspace{\parindent}
The displacement field is approximated by a deep feedforward neural network, i.e. $\mathbf{u} \cong \mathcal{N}:\Omega_{m}\times H_{\lambda}\longrightarrow \mathbb{R}^{2}$, where:
\begin{equation} \label{eq:221e1}
\begin{aligned}
\mathcal{N}: (\mathbf{x};\boldsymbol{\lambda}) \longmapsto g^{N} \circ g^{N-1} \circ \cdots \circ g^{1}(\mathbf{x})
\end{aligned}
\end{equation}
where $\boldsymbol{\lambda} = [\mathbf{W},\mathbf{b} ]  \in H_{\lambda}$ denotes the trainable parameters $\mathbf{W}$ and $\mathbf{b}$, while $g^{i} = \sigma^{i}(\mathbf{W}^{i} \cdot \mathbf{z}^{i} + \mathbf{b}^{i})$ denotes the $i^{th}$ layer maps of the $N^{th}$ neural network layer, respectively, with $\sigma^{i}$ being the corresponding activation function, whereas $\mathbf{z}^{i}$ denotes the layer inputs. The PINN framework requires the addition of information arising from the physical model to the loss function. More specifically, the output of the neural network is forced to satisfy the partial differential operator of a specific physics problem on a selected set of points that are interior to the domain of interest \cite{RAISSI2019686,RAISSI2018125,712178}. Hence, the loss function $\mathcal{L}(\boldsymbol{\lambda})$ is expressed as the weighted sum of the loss term concerning the physics $\mathcal{E}$, the loss terms of the boundary $\mathcal{B}$, and domain data $\mathcal{D}$, as follows:
\begin{equation} \label{eq:221e2}
\begin{aligned}
\mathcal{L}(\boldsymbol{\lambda}) = w_{1}\mathcal{E}(\boldsymbol{\lambda}) + w_{2}\mathcal{B}(\boldsymbol{\lambda}) + w_{3}\mathcal{D}(\boldsymbol{\lambda})
\end{aligned}
\end{equation}
where the weights $w_{1},w_{2},w_{3}$ range between $[0,1]$ and are adjusted accordingly to provide a different weight to each of the loss terms. 

In a Variational Physics Informed Neural Network, instead of the differential operator, the weak formulation of the physical problem is forced to be satisfied \cite{KHARAZMI2021113547,pmlr-v120-khodayi-mehr20a}. Based on Eq. \eqref{eq:21e4b} and considering a set of $N_{w}$ test functions $\mathbf{w}^{k} \in W$ ($k=1,2,...,N_{w}$), the physics-based loss term $\mathcal{E}(\boldsymbol{\lambda})$ takes the form:
\begin{equation} \label{eq:221e4}
\begin{aligned}
&\mathcal{E}(\boldsymbol{\lambda}) = \sum_{\Omega_{m}} \sum_{k=1}^{N_{w}}
\Big (   
\int_{\Omega_{m}} \nabla \mathbf{w}^{k} : \boldsymbol{\Sigma}(\mathcal{N}(\mathbf{x};\boldsymbol{\lambda})) d\Omega - 
\int_{\partial\Omega_{m}} \mathbf{w}^{k} \cdot \mathbf{t}(\mathcal{N}(\mathbf{x};\boldsymbol{\lambda})) d\Gamma
\Big ) ^{2}
\end{aligned}
\end{equation}
Here the neural network $\mathcal{N}$ approximates the weak PDE solution $\mathbf{u}$, therefore, an appropriate solution Hilbert space for Eq. \eqref{eq:21e4b} is $V = \{ \mathcal{N}:\Omega_{m}\times H_{\lambda}\rightarrow \mathbb{R}^{2} \} \cap H^{1}(\Omega_{0})$ \cite{rojas2024}.

The choice of the test function $\mathbf{w}$ leads to a specific variational formulation \cite{KHARAZMI2021113547}. Focusing on a local Petrov-Galerkin formulation, for each domain $\Omega_{m}$, a discretization $\mathcal{T}(\Omega_{m})$ is required, whereas for each subdomain $E \in \mathcal{T}(\Omega_{m})$, the test function $\mathbf{w}_{E}^{k}$ is defined as: $\mathbf{w}_{E}^{k} = [\Psi_{E}^{k}(\boldsymbol{\xi}), \Psi_{E}^{k}(\boldsymbol{\xi})]^{T}$. For the present study, the shape functions of the linear 4-node element are employed as test functions, therefore, $N_{w} = 4$ and $\Psi_{E}^{k}(\boldsymbol{\xi})$ denotes the shape function that corresponds to the $k^{th}$ node of the subdomain $E$. It should be noted that in the case of the strong form PINN formulation, the loss term $\mathcal{E}$ of Eq. \eqref{eq:221e2} concerns the Navier differential operator:
\begin{equation} \label{eq:221e4b}
\begin{aligned}
&\mathcal{E}(\boldsymbol{\lambda}) = \frac{1}{\#P_{\Omega}} \sum_{i=1}^{\#P_{\Omega}} \big \| \upmu \nabla^{2}\mathcal{N}(\mathbf{x}^{i};\boldsymbol{\lambda}) + (\uplambda + \upmu) \nabla \nabla \cdot \mathcal{N}(\mathbf{x}^{i};\boldsymbol{\lambda})  \big \|^{2}
\end{aligned}
\end{equation}
where $P_{\Omega}$ in Eq. \eqref{eq:221e4b} is a set of points $\mathbf{x} \in \Omega_{0}$, that are used for the collocation of the Navier's differential operator. In order to define the remaining loss terms of Eq. \eqref{eq:221e2}, a set of points $S_{\partial \Omega}$ is considered, located on the boundary $\partial \Omega_{0} \cup \Gamma_{I}$ where the target values of the displacement field are given. Moreover, a set of additional auxiliary points $S_{\Omega}$, interior to the $\Omega_{0}$ domain, with known displacement vector is considered. As such, the following training sets $D_{\partial\Omega} = \{ (\mathbf{x}^{i}, \mathbf{u}^{i}) | \mathbf{x}^{i} \in S_{\partial \Omega}, \hspace{0.1cm} \mathbf{u} = \mathbf{u}(\mathbf{x}^{i})  \}$ and domain data $D_{\Omega} = \{ (\mathbf{x}^{i}, \mathbf{u}^{i}) | \mathbf{x}^{i} \in S_{\Omega}, \hspace{0.1cm} \mathbf{u} = \mathbf{u}(\mathbf{x}^{i})  \}$ are defined. Given the aforementioned training sets, the present work employs the Mean Squared Error (MSE) expressions for the loss terms $\mathcal{B}$ and $\mathcal{D}$, i.e.:
\begin{equation} \label{eq:221_mse}
\begin{aligned}
&\mathcal{B}(\boldsymbol{\lambda}) = \frac{1}{\#D_{\partial\Omega}} \sum_{i=1}^{\#D_{\partial\Omega}} \big \|\mathcal{N}(\mathbf{x}^{i};\boldsymbol{\lambda}) - \mathbf{u}^{i} \big \|^{2}
, \hspace{0.5cm}
&\mathcal{D}(\boldsymbol{\lambda}) = \frac{1}{\#D_{\Omega}} \sum_{i=1}^{\#D_{\Omega}} \big \| \mathcal{N}(\mathbf{x}^{i};\boldsymbol{\lambda}) - \mathbf{u}^{i} \big \|^{2}
\end{aligned}
\end{equation}

The training process aims to find the parameters $\boldsymbol{\lambda}^{*}$ such that:
\begin{equation} \label{eq:221e5}
\begin{aligned}
\boldsymbol{\lambda}^{*} = \argminA_{\boldsymbol{\lambda}} \mathcal{L}(\boldsymbol{\lambda})
\end{aligned}
\end{equation}

The minimization problem \eqref{eq:221e5} is solved iteratively through a gradient-based procedure, employing the Adam (Adaptive Moment Estimation) optimization algorithm \cite{DBLP:journals/corr/KingmaB14}. The derivatives of Eq. \eqref{eq:221e4} with respect to the position vector $\mathbf{x}$, as well as the gradient components of the loss function (Eq. \eqref{eq:221e2}) with respect to the network parameters $\boldsymbol{\lambda}$ are computed by implementing automatic differentiation \cite{JMLR:v18:17-468}, implemented in Tensorflow \cite{tensorflow2015-whitepaper}.
Figure \ref{fig:training} summarizes the training process of the VPINN. The neural network takes as input the position vector that belongs to the subdomains $\Omega_{1},\Omega_{2}$ or on their boundaries $\partial\Omega_{1},\partial\Omega_{2}$, while the displacement vector field is calculated as the network output. Then, for the interior integration points, the stress tensor is calculated, evaluating the derivatives of the displacement components.
\begin{figure}[H]
\centering
\includegraphics[scale=0.4]{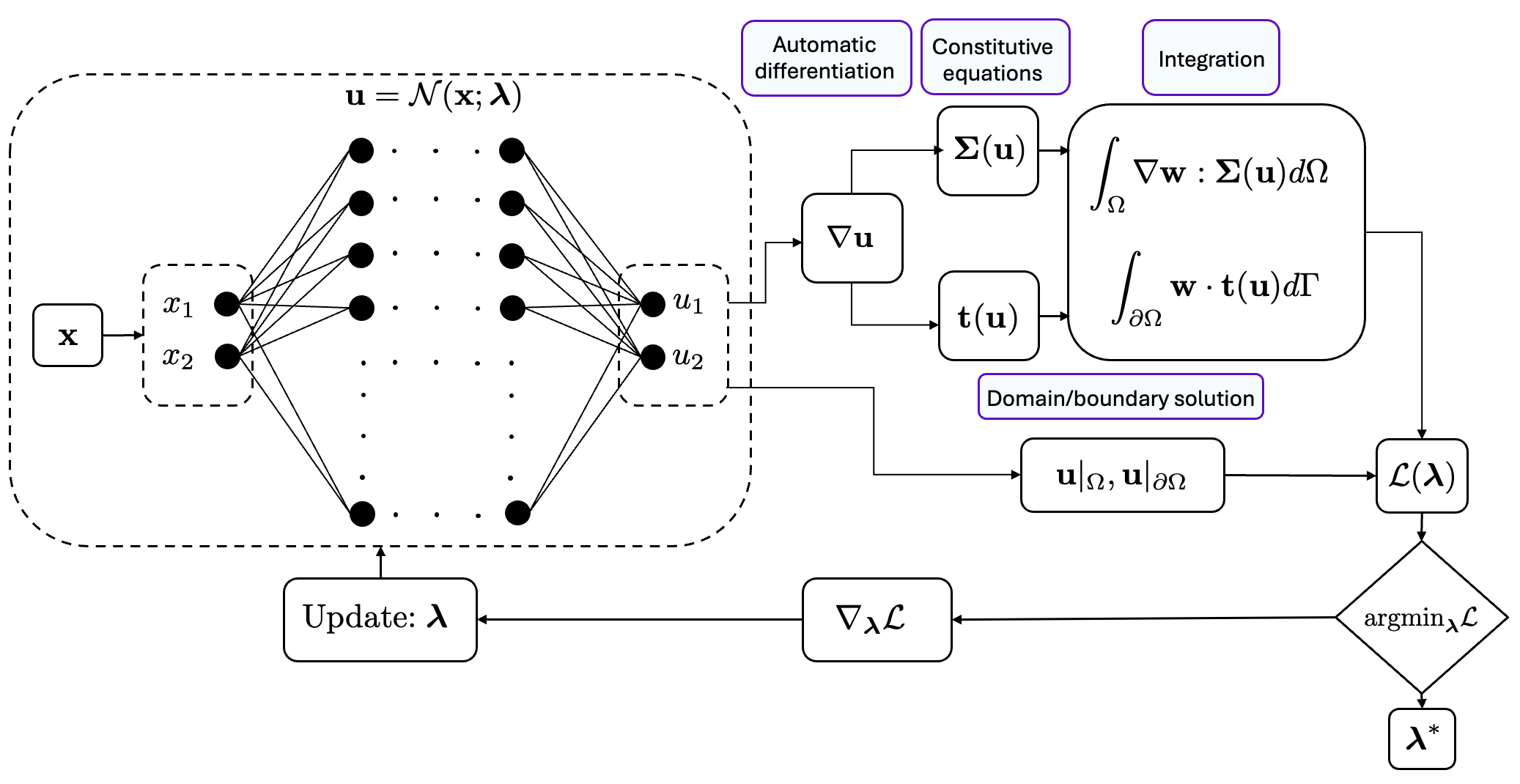} 
\caption{The training process of the Neural Network based on a Variational formulation.}
\label{fig:training}
\end{figure}

\subsubsection{Tree-based integration}
\label{Sec:tree}
\hspace{\parindent}
An indispensable part of the formulation is the evaluation of the weak-form integrals of Eq. \eqref{eq:221e4}. In the present work, a tree-based integration algorithm is developed for the numerical integration of the weak formulation. The starting point of the algorirthm concerns the segmentation of the subdomains $\Omega_{1}$ and $\Omega_{2}$ with the aid of a quadtree $\mathcal{Q}(\Omega_{0})$, as depicted in Fig. \ref{fig:integration}a. Considering the cells $E$ as the vertices of the quadtree $\mathcal{Q}(\Omega_{0})$, the set $S(E)$ is formed by the sons $\tau$ of each vertex cell $E$, containing $4$ square cells of half size. The segmentation is obtained recursively as follows:
\begin{enumerate}
    \item Setting the bounding box $E = r$ as the root cell, i.e.: $r \subseteq \Omega_{0}$, the level $0$ ($l_{0}$) of the quadtree is defined. 
    \item A set of points $\boldsymbol{\chi}$ is distributed uniformly in $E$. Then, the following sets are defined: $\Xi_{1} = \{ \boldsymbol{\chi} \in E | E \cap \Omega_{1} \neq \emptyset \}$ and $\Xi_{2} = \{ \boldsymbol{\chi} \in E | E \cap \Omega_{2} \neq \emptyset \}$
    \item The following splitting criterion is defined: If $\Xi_{1} \neq \emptyset$ and $\Xi_{2} \neq \emptyset$ then split the cell $E=r$: $E \mapsto S(E)$. For each of the sons $\tau \in S(E)$ repeat the steps 2 and 3. If the splitting criterion is not satisfied then the set of sons is empty $S(E) = \emptyset$.  
    \item Upon completing the segmentation process, the set of the tree leafs is defined as $\mathcal{T}(\Omega_{0}) = \{ E \in \mathcal{Q}(\Omega_{0}) | S(E) = \emptyset \}$. The set of integration cells is defined as the disjoint set $\mathcal{T}(\Omega_{1}) \sqcup \mathcal{T}(\Omega_{2}) \subset \mathcal{T}(\Omega_{0})$ containing the leafs $\{E \in \mathcal{T}(\Omega_{0}) | E \cap \Omega_{1} = \emptyset, E \cap \Omega_{2} = \emptyset\}$.
\end{enumerate}

For each cell $E$, a quadrature rule $\mathcal{G}(E) = \{ (\boldsymbol{\xi}, h(\boldsymbol{\xi})) \}$ is employed to generate the integration points $\boldsymbol{\xi}$ and weights $h(\boldsymbol{\xi})$. Thus, the sets $\mathcal{G}(\mathcal{T}(\Omega_{1})) = \bigcup_{E \in \mathcal{T}(\Omega_{1})} \mathcal{G}(E)$ and $\mathcal{G}(\mathcal{T}(\Omega_{2})) = \bigcup_{E \in \mathcal{T}(\Omega_{2})} \mathcal{G}(E)$ are formed, containing the pairs $(\boldsymbol{\xi}, h(\boldsymbol{\xi})) \in \mathcal{G}(E)$ for all $E \in \mathcal{T}(\Omega_{1}) \sqcup \mathcal{T}(\Omega_{2})$, as shown in Fig. \ref{fig:integration}b. Mapping the pair of each cell $(\boldsymbol{\xi}, h(\boldsymbol{\xi}))$ to the global sets of integration points $\mathcal{G}(\mathcal{T}(\Omega_{1}))$, $\mathcal{G}(\mathcal{T}(\Omega_{2}))$, the integration of a function $f: \Omega \longrightarrow \mathbb{R}$ can be represented as the following matrix multiplication:
\begin{equation} \label{eq:222e1}
\begin{aligned}
\int_{\Omega}f d\Omega \cong \sum_{E\in \mathcal{T}(\Omega)} \int_{E} f d\Omega \cong \sum_{E\in \mathcal{T}(\Omega)} \sum_{(\boldsymbol{\xi},h(\boldsymbol{\xi})) \in \mathcal{G}(E)} h(\xi)f(\xi) = \sum_{(\boldsymbol{\xi},h(\boldsymbol{\xi})) \in \mathcal{G}(\Omega)} h(\boldsymbol{\xi})f(\boldsymbol{\xi}) = \mathbf{h}^{T} \cdot \mathbf{f}
\end{aligned}
\end{equation}
\noindent where the vectors $\mathbf{f}$ and $\mathbf{h}$ contain the function evaluations $f(\xi)$ on the integration points and the corresponding weights $h(\xi)$. Therefore, the weak form integral of Eq. \eqref{eq:221e4} is evaluated as:
\begin{equation} \label{eq:222e2}
\begin{aligned}
\int_{\Omega_{m}} \nabla \mathbf{w}^{k} : \boldsymbol{\Sigma}(\mathcal{N}(\mathbf{x};\boldsymbol{\lambda})) d\Omega 
&\cong
\sum_{E\in \mathcal{T}(\Omega)}
\int_{E} \nabla \mathbf{w}_{E}^{k} : \boldsymbol{\Sigma}(\mathcal{N}(\mathbf{x};\boldsymbol{\lambda})) d\Omega = \sum_{(\boldsymbol{\xi},h(\boldsymbol{\xi})) \in \mathcal{G}(\Omega)} h(\boldsymbol{\xi})f(\boldsymbol{\xi}) = \mathbf{h}^{T} \cdot \mathbf{f}
\\
&f(\boldsymbol{\xi}) = \nabla \mathbf{w}_{E}^{k}(\boldsymbol{\xi}) : \boldsymbol{\Sigma}(\mathcal{N}(\boldsymbol{\xi};\boldsymbol{\lambda}))
\end{aligned}
\end{equation}

The adaptive sizing of the integration cells due to the quadtree structures offers the advantage of computational efficiency during the integration procedure, as smaller cells are used near the domain boundaries, where refinement is required, while a coarser distribution of cells is used for the rest of the domain. Furthermore, the generation of quadtree representations requires only boundary information, providing substantial flexibility in the treatment of different multi-phase materials with complex inner architetctures, without the use of extra CAD tools.
\begin{figure}[H]
\centering
\includegraphics[scale=0.55]{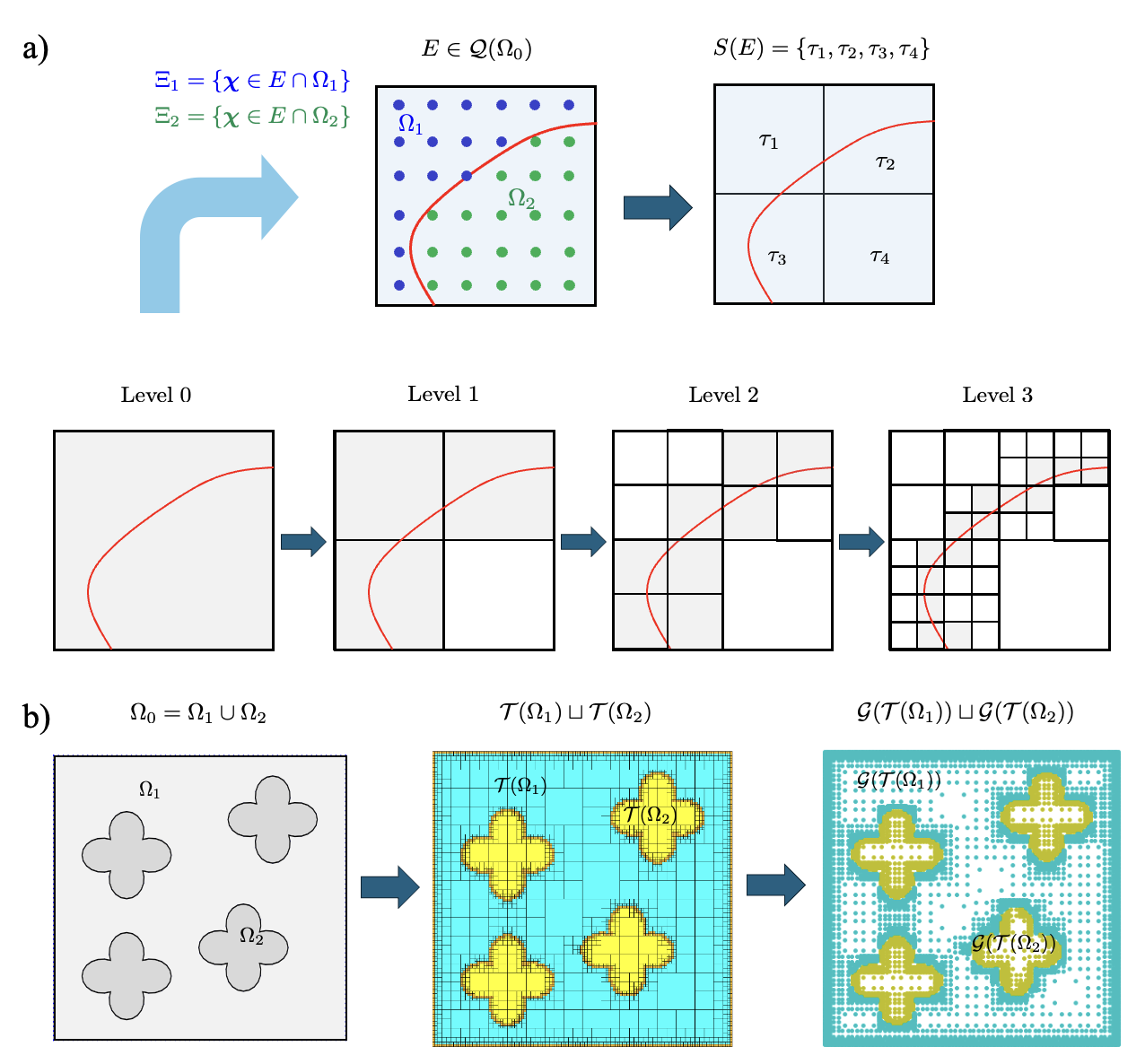} 
\caption{Segmentation of $\Omega_{0} = \Omega_{1} \cup \Omega_{2}$. The subdomains $\Omega_{1}$ and $\Omega_{2}$ are discretized using the quadtree representation $\mathcal{Q}(\Omega_{0})$. a) The splitting procedure of a single cell of the quadtree $\mathcal{Q}(\Omega_{0})$. b) The formation of the sets $\mathcal{T}(\Omega_{1})$ and $\mathcal{T}(\Omega_{2})$. The integration cells are the leafs $E \in \mathcal{T}(\Omega_{1}) \sqcup \mathcal{T}(\Omega_{2})$. Employing a quadrature rule $\mathcal{G}(E)$ for each subcell $E$ the sets of integration points $\mathcal{G}(\mathcal{T}(\Omega_{1})) \sqcup \mathcal{G}(\mathcal{T}(\Omega_{2}))$ are formed.}
\label{fig:integration}
\end{figure}

\subsection{Machine Learning Framework}
\label{Sec:ML_Frame}
\subsubsection{Galerkin Boundary Element computation of boundary and inner domain data}
\hspace{\parindent}
The elastic boundary value problem \eqref{eq:21e1}-\eqref{eq:21e3} can be reformulated as a set of integral equations \cite{BEMSolids}:
\begin{equation}  \label{eq:231e1}
\begin{aligned}
\big(\mathcal{U}_{\Omega_{1}}\big)(\mathbf{x}') &= c(\mathbf{x}')\mathbf{u}^{1}(\mathbf{x}')+
\int_{\partial\Omega_{0} \cup \Gamma^{1}_{I}} \Tilde{\mathbf{T}}(\mathbf{x},\mathbf{x}') \cdot\mathbf{u}^{1}(\mathbf{x}) d\Gamma
- 
\int_{\partial\Omega_{0} \cup \Gamma^{1}_{I}}  \Tilde{\mathbf{U}}(\mathbf{x},\mathbf{x}')   \cdot\mathbf{t}^{1}(\mathbf{x}) d\Gamma = \mathbf{0}
\\
\big(\mathcal{U}_{\Omega_{2}}\big)(\mathbf{x}') &= (1 - c(\mathbf{x}'))\mathbf{u}^{2}(\mathbf{x}')+
\int_{\Gamma^{2}_{I}} \Tilde{\mathbf{T}}(\mathbf{x},\mathbf{x}') \cdot\mathbf{u}^{2}(\mathbf{x}) d\Gamma
- 
\int_{\Gamma^{2}_{I}}  \Tilde{\mathbf{U}}(\mathbf{x},\mathbf{x}')   \cdot\mathbf{t}^{2}(\mathbf{x}) d\Gamma = \mathbf{0}
\end{aligned}
\end{equation}
\begin{figure}[H]
\centering
\includegraphics[scale=0.7]{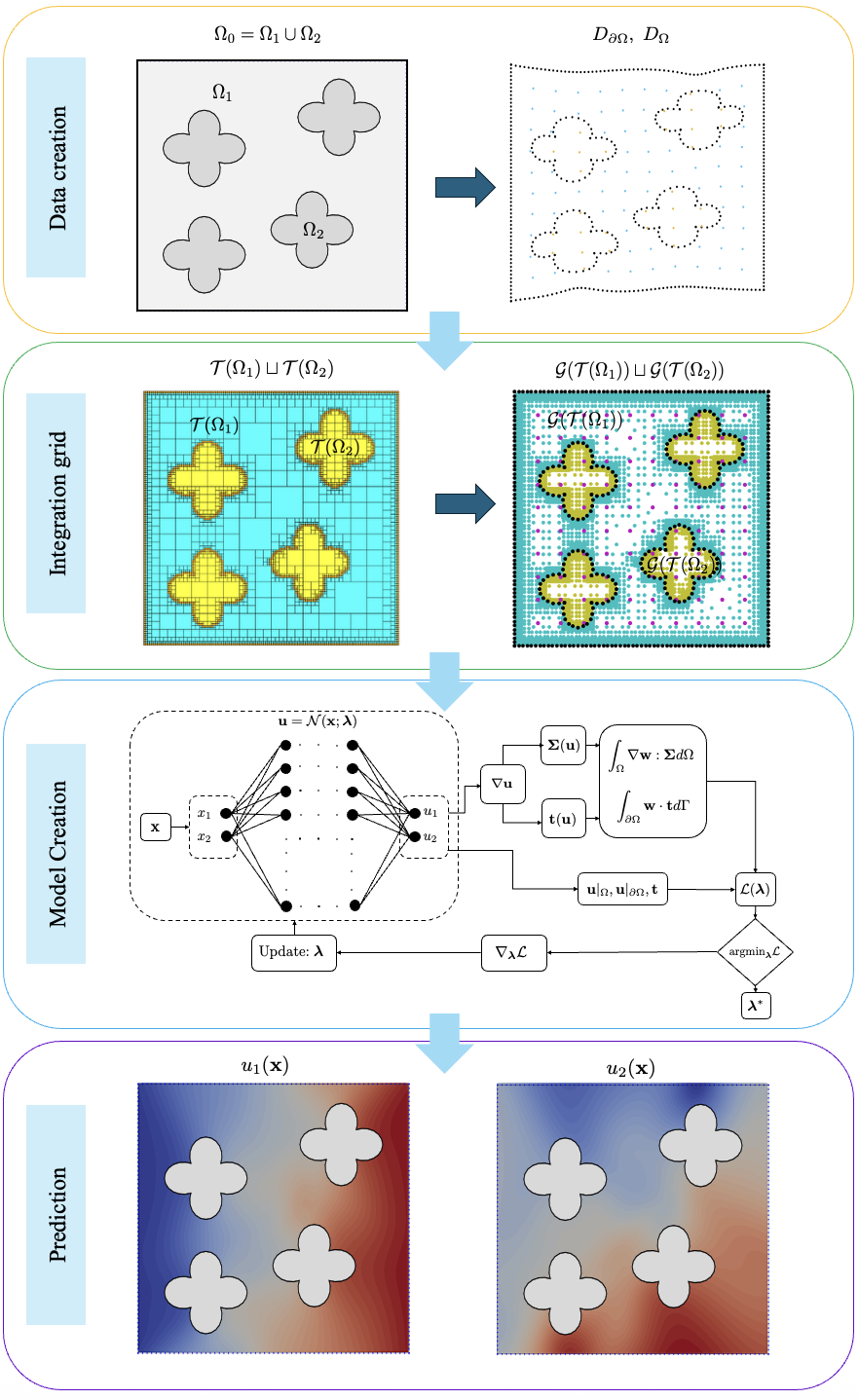} 
\caption{The proposed Machine learning framework. The first step is the data generation through the GBEM scheme. Using the information of the domain boundaries, the integration grid is created via quadtree representations. Then, the VPINN model is trained and utilized for predictions.}
\label{fig:framework}
\end{figure}
where $\mathbf{x}$, $\mathbf{x}'$ denote the source and field points, respectively, whereas the jump coefficient $c(\mathbf{x}')$ obtains the values $\frac{1}{2}$, $1$, and $0$ for $\mathbf{x}' \in \Gamma$, $\mathbf{x}' \in \Omega_{m}$  and $\mathbf{x}' \notin \Omega_{m}$, respectively. The integral kernels $\Tilde{\mathbf{U}}(\mathbf{x},\mathbf{x}')$, $\Tilde{\mathbf{T}}(\mathbf{x},\mathbf{x}')$ entering Eq. \eqref{eq:231e1} are provided in \cite{RODOPOULOS2024112603}. For the analysis, a weak formulation is employed, multiplying the integral equations \eqref{eq:231e1} with a vector weight function $\boldsymbol{\upsilon}(\mathbf{x}') \in \Sigma_{m}$, respectively, where $\Sigma_{m}=[L^{2}(\Gamma_{m})]^{2}$, with $L^{2}(\Gamma_{m})$ denote the space of square-integrable functions \cite{evans10} defined on $\Gamma_{m}$.  Integrating over the boundaries $\Gamma_{1}$ and $\Gamma_{2}$ with respect to the coordinates of $\mathbf{x}'$ \cite{Sutradhar2008}, the following weak formulation is obtained \cite{Sutradhar2008}:
\begin{equation}  \label{eq:231e2}
\begin{aligned}
\text{Find} \hspace{0.1cm} \mathbf{u}^{m} \in \Sigma_{m} \hspace{0.1cm} \text{such that} \hspace{0.1cm} \int_{\Gamma_{m}} \boldsymbol{\upsilon}(\mathbf{x}') \cdot \big(\mathcal{U}_{\Omega_{m}}\big)(\mathbf{x}') d\Gamma' = 0 \hspace{0.1cm} \text{for all} \hspace{0.1cm} \mathbf{w}^{m} \in \Sigma_{m}
\end{aligned}
\end{equation}

Following the standard discretization procedure detailed in \cite{RODOPOULOS2025113159} and taking into account the corresponding boundary conditions Eq. \eqref{eq:21e3}, the following set of matrix equations is obtained:
\begin{equation} \label{eq:231e3}
\begin{aligned}
\Tilde{\mathbf{A}} \cdot \hat{\mathbf{X}} &= \hat{\mathbf{Y}}
\end{aligned}
\end{equation}
where $\hat{\mathbf{X}}$ stands for the vector that contains all the unknown nodal values of the displacement and traction vectors, while the left-hand matrix $\Tilde{\mathbf{A}}$ and the right-hand vector $\hat{\mathbf{Y}}$ contain integrals (Eq. \eqref{eq:231e2}), evaluated numerically according to \cite{RODOPOULOS2024112603}. By solving the algebraic system \eqref{eq:231e3}, the displacement field can be accurately calculated at any point $\mathbf{x}'$ interior to $\Omega_{1}$ and $\Omega_{2}$, using the integral equations \eqref{eq:231e1}.

\subsubsection{Training and performance assessment}
\hspace{\parindent}
Defining the discretization of the exterior and interior boundary of $\Omega_{0}$, the aforementioned GBEM is employed for the data generation, used for the machine learning formulation. To this point, the nodes of the boundary element mesh, along with calculated displacement values by the solution of Eq. \eqref{eq:231e3}, form the boundary training set $D_{\partial \Omega}$. Furthermore, selected auxiliary points that are located interior to $\Omega_{0}$ define the set $S_{\Omega}$. The displacement vector on these points is obtained semi-analytically employing the integral equations (Eq. \eqref{eq:231e1}), adjusting the values of the jump coefficient $c(\mathbf{x}')$ appropriately ($c(\mathbf{x}') = 1$ if $\mathbf{x}' \in \Omega_{1}$ or $c(\mathbf{x}') = 0$ if $\mathbf{x}' \in \Omega_{2}$). By that means, the auxiliary training set $D_{\Omega}$ is formed.

The next step is the discretization of $\Omega_{0}$ into cells, as required for the computation of the weak form integrals of Eq. \eqref{eq:221e4}. This is achieved via the segmentation of $\Omega_{0}$ employing a cluster quadtree, as described in \ref{Sec:tree}, therefore, defining the integration cell sets $\mathcal{T}(\Omega_{1})$ and $\mathcal{T}(\Omega_{2})$. For the numerical integration of Eq. \eqref{eq:221e4} a Gauss-Legendre point grid $\mathcal{G}(E)$ is defined in each integration cell $E$. The deep NN (Eq. \eqref{eq:221e1}) is subsequently trained employing $N_{l}$ hidden layers, containing $N_{n}$ neurons. The deep learning model calculates the displacement vector so that it satisfies the solid mechanics boundary value problem at every domain point $\mathbf{x} \in \Omega_{0}$ (Section \ref{Sec:vpinn}). In each training epoch, the numerical integrations of the network output, as well as the data $D_{\partial \Omega}$, $D_{\Omega}$ are utilized to compute the terms $\mathcal{E}(\boldsymbol{\lambda})$, $\mathcal{B}(\boldsymbol{\lambda})$ and $\mathcal{D}(\boldsymbol{\lambda})$, entering the loss function definition of Eq. \eqref{eq:221e2}. Upon completion of the training procedure, the network is employed to predict the displacement vector $\mathbf{u}$ at points in $\Omega_{0}$ that are not seen during the training procedure, forming the testing set $T_{\Omega}$.

The accuracy of the numerical predictions of the VPINN formulation with respect to the given testing set $T_{\Omega}$ is assessed through the relative error $e(u; u_{t})$ and the coefficient of determination $R^{2}(u; u_{t})$, defined as follows:
\begin{equation} \label{eq:232e1}
\begin{aligned}
&e(u; u_{t})=\frac{\| u - u_{t}\|_{2}}{\| u_{t}\|_{2}}, \hspace{0.8cm} 
R^{2}(u; u_{t})= 1 - \Bigg (\frac{\| u - u_{t}\|_{2}}{\| u_{t} - \bar{u}_{t}\|_{2}} \Bigg)^{2}
\\
& \text{with} \hspace{0.2cm} \| u\|_{2} = \Big( \int_{\Omega_{0}}u^{2} d\Omega \Big)^{\frac{1}{2}}
\end{aligned}
\end{equation}
where $u$ and $u_{t}$ represent each component of the predicted and true solution, respectively, $e(u; u_{t})$ denotes the relative error with respect to the true solution (VFEM calculations), provided in the testing data $(\mathbf{x},\mathbf{u}_{t}) \in T_{\Omega}$.

}


\section{Results} 
\label{Sec:Results}
{
\subsection{Phase patterns and hyperparameter settings} 
\hspace{\parindent}
We investigate the performance of two-dimensional double-phase materials, where the second phase $\Omega_{2}^{i}$ topology follows the Gielis' formula \cite{RODOPOULOS2025113159,Matsuura2015}:
\begin{equation} \label{eq:3a}
\begin{aligned}
\mathbf{x}(\varphi) = S(|\cos{\frac{m}{4}\varphi}|+|\sin{\frac{m}{4}\varphi}|)^{n}\mathbf{R}(\beta) \cdot
\left \{
\begin{matrix}
\cos{\varphi}
    \\
\sin{\varphi}
\end{matrix}
\right \}
\end{aligned}
\end{equation}
where $\mathbf{x}(\varphi)$ in Eq. \eqref{eq:3a} denotes every point located on the curve that defines the boundary of $\Omega_{2}^{i}$ with $\Omega_{1}^{i}$. The boundary is formed by the evolution of the parameter $\varphi \in [0,2\pi)$, while $m$ and $n$ are real-valued constants that control its shape. Moreover, $S$ is a magnification parameter and $\mathbf{R}(\beta)$ is the rotation matrix that defines the angular rotation of the second-phase by an angle $\beta$. Among the geometry parameters of Eq. \eqref{eq:3a}, the constant $n$ is the main parameter that controls the shape of the second phase ($n<0$: star-shaped, $n>0$: flower-shaped, $n=0$: circular), while the parameter $m$ is the number of repetitions of an open curve which forms the second phase \cite{RODOPOULOS2025113159}.

Four numerical test cases with different second-phase arrangements and geometries are considered, with a volume fraction $V_{f} = 20\%$. An external material $\Omega_{1}$ with a Young's modulus $E^1 = 1$ and Poisson's ratio $\nu^1 = 0.3$ is prescribed, with a second phase material $\Omega_{2}$ having an elastic modulus $E^2 = 0.1$ and $\nu^2 = 0.3$. In all cases, the same deep NN with $N_{l} = 5$ layers and $N_{n} = 60$ neurons per layer is employed. The accuracy of the numerical predictions is evaluated by testing different combinations of hyperparameters regarding the network architecture (number of layers, number of neurons) through a sensitivity analysis performed and presented in detail in \nameref{appA}. For the numerical integration of the weak form integrals of Eq. \eqref{eq:221e4}, a quadtree with $7$ levels is constructed to provide the set of integration cells, while a $2\times2$ grid of Gauss-Legendre points is considered for each cell. For the training procedure, a fixed number of 20000 epochs is used, with a variable learning rate that starts from $r_{l} = 0.001$ and decreases to half, after 3000 epochs. 

Two different models are considered for each test case. The first model is based solely on the use of the boundary displacement data $D_{\partial \Omega}$, while the weak form is enforced to be satisfied in the interior of $\Omega_{1}$ and $\Omega_{2}$. This is achieved by setting the weighting coefficients of Eq. \eqref{eq:221e2} as $(w_{1},w_{2},w_{3})=(1,1,0)$. For the second model, the effect of additional points $D_{\Omega}$ inside $\Omega_{1}$ and $\Omega_{2}$ with known target values of the displacement vector (GBEM formulation) is evaluated by setting $(w_{1},w_{2},w_{3})=(1,1,1)$. The predicted results are compared with the numerical results obtained by the Voxel Finite Element Method (VFEM) \cite{Bez2024}. The VFEM scheme employs a refined discretization pattern of 80$\times$80, resulting in a total of $6400$ quadratic volume elements. The nodes of the VFEM mesh are used as unseen data to the trained deep learning model. All the training and simulation procedures have been implemented on the HPC 2.24 PFLOPs (33,984 CPU cores) cluster hosted at NYUAD Data Center. For the computational tasks, a single cluster AMD EPYC Rome CPU node with 128 cores and 512GB RAM has been used. 
\begin{figure}[H]
\centering
\includegraphics[scale=0.55]{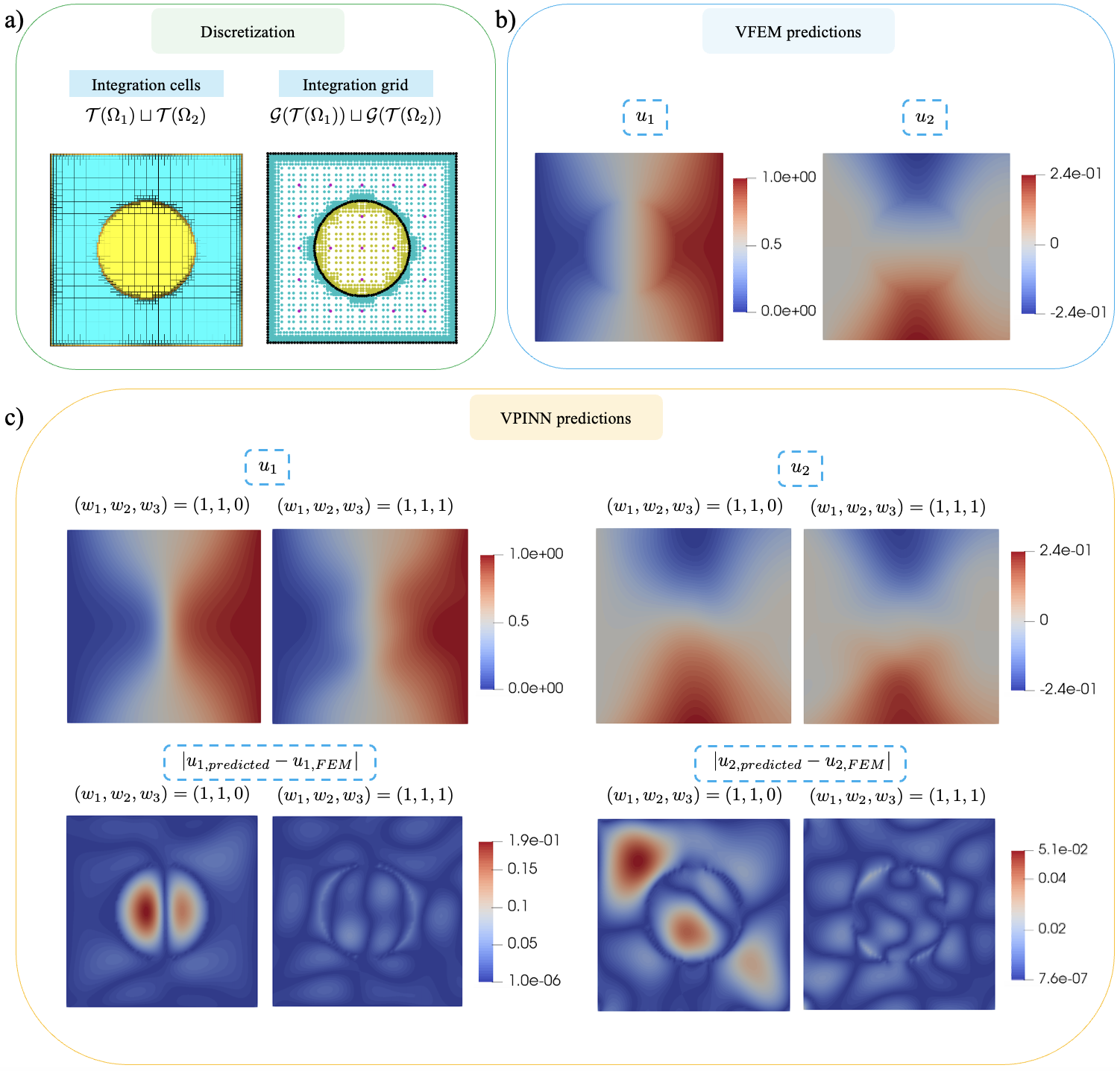} 
\caption{Plane-stress plate with one circular inclusion ($n=0$): a) Integration cells and training point grid, b) Numerical results for the displacement components acquired from the VFEM formulation, and, c) Contour plots of the displacement prediction of the VPINN formulation with ($w_{3}=1$) and without ($w_{3}=0$) taking into account target values on the additional auxiliary interior points. The contour plots of the point-wise absolute errors are provided as well using a common scale in the colormap.}
\label{fig:fig5}
\end{figure}

\subsection{Single circular second-phase}
\label{sec:Circ_SP}

A plane stress plate $[0,1] \times [0,1]$ with a circular second phase ($n=0$, Eq. \eqref{eq:3a}) is depicted in Figure \ref{fig:fig5}a with the set of 2224 integration cells defined as the set of leaves of the constructed quadtree. The integration points (light blue and yellow points) are created by imposing a $2\times2$ Gaussian point grid in each integration cell. The training boundary displacement data is generated using the GBEM scheme, with a surface mesh of 274 quadratic boundary elements (black dots). Additional displacement data are provided on a grid of $5 \times 5$ interior points (purple points in Fig. \ref{fig:fig5}a), using the integral equations (Eq. \eqref{eq:231e1}). 

\begin{figure}[H]
\centering
\includegraphics[scale=0.4]{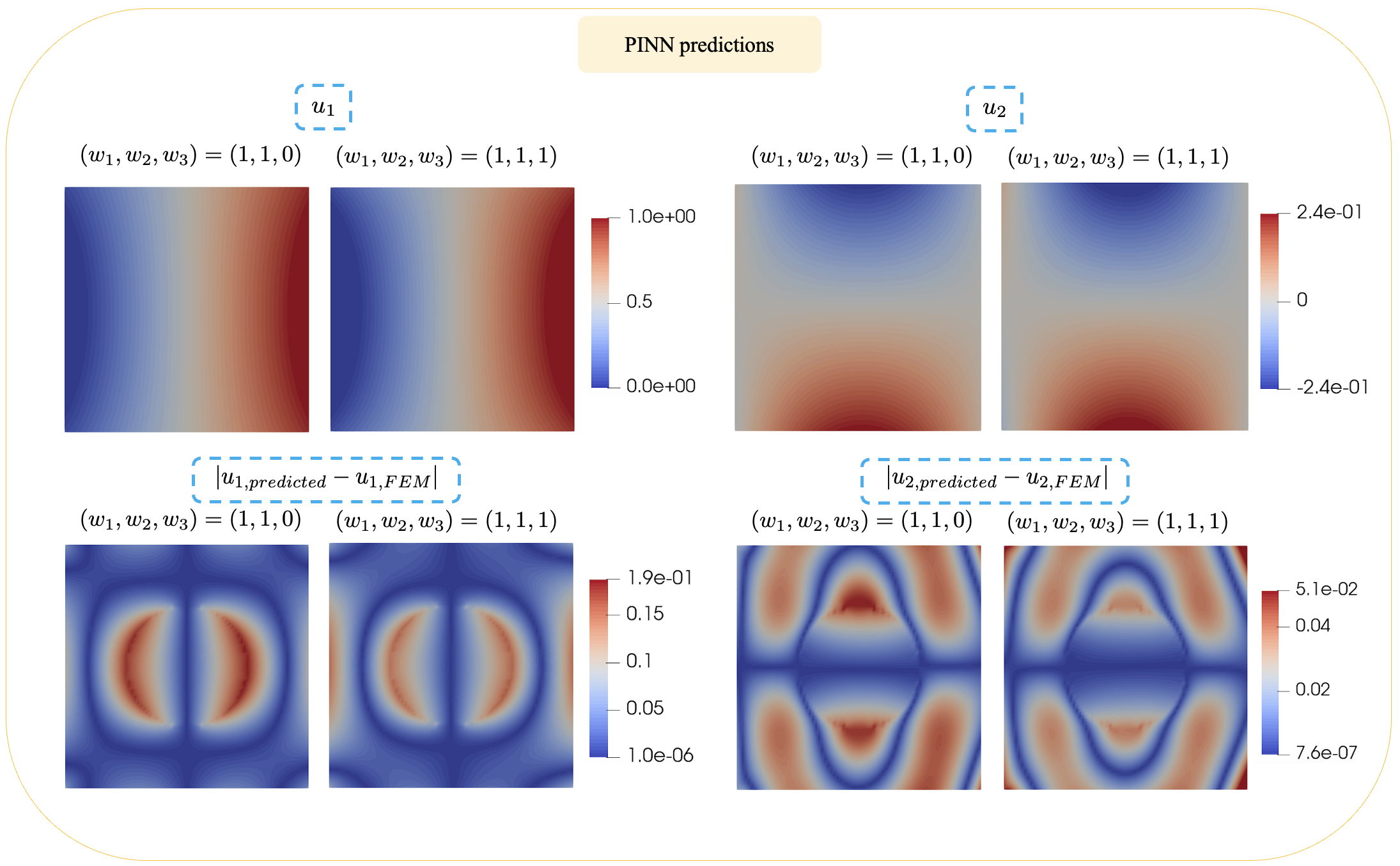} 
\caption{Plane-stress plate with one circular inclusion ($n=0$): Contour plots of the displacement prediction of the classical PINN formulation with ($w_{3}=1$) and without ($w_{3}=0$) taking into account target values on the additional auxiliary interior points. The contour plots of the point-wise absolute errors are provided as well using a common scale in the colormap.}
\label{fig:fig5b}
\end{figure}
Fig. \ref{fig:losses}a provides the evolution of the normalized loss function $\mathcal{L}(\boldsymbol{\lambda})/ \mathcal{L}_{0}$ during the training procedure, where $\mathcal{L}_{0}$ denotes the loss value of the first epoch. The contour plots of the displacement components $u_{1}$ and $u_{2}$ obtained from the VFEM formulation are presented in Fig. \ref{fig:fig5}b. Figure \ref{fig:fig5}c provides the contour plots of the predicted results for $u_{1}$ and $u_{2}$, in the VFEM result scale, using the VPINN formulation for two different loss function cases, namely for $w_{3}=0$ and $w_{3}=1$, thus without and with the internal point data information.


The results of the deep learning VPINN model with $w_{3}=0$ capture the field distribution of $\Omega_{1}$ with considerable accuracy, but fail to capture the displacement field of the second phase $\Omega_{2}$, as indicated by the corresponding error magnitudes. The use of additional inner data ($w_{3}=1$) for the same deep learning model substantially improves the accuracy of the VPINN predictions for the same number of training epochs. This can be verified from the point-wise absolute errors in Fig. \ref{fig:fig5}c, where the values for both components are significantly reduced by including the additional interior point data.

For the specific test case, classical strong form PINN results are provided for comparison purposes. In this case, a grid of 3080 unimorphically distributed points, interior to $\Omega_{0}$, is used for the collocation of the Navier's differential operator of Eq. \eqref{eq:221e4b}. The contour plots of the numerical predictions, as well as the pointwise errors of the PINN framework, are shown in Fig. \ref{fig:fig5b}. It can be observed that, for the same NN architecture and the same training setup (optimizer settings, training epochs), the strong form PINN scheme fails to capture the discontinuity caused by the existence of the second material phase, for both Loss function definitions, without and with inner domain data ($w_{3} = 0$ and $w_{3} = 1$). The displacement components $u_{1}$, $u_{2}$ (Fig. \ref{fig:fig5b}) reveal a smooth field distribution, as in the case of a uniform and homogeneous material. The quality of the results is reflected in the point errors, where intense error values are concentrated in the neighborhood of the interface boundary $\Gamma_{I}$. The addition of internal point information in the VPINN case (Fig. \ref{fig:fig5}) substantially improves the accuracy of the displacement field, reducing the maximum displacement field differences below 10$\%$. This is not the case for the classical PINN formulation, as it can be observed from the corresponding displacement field error magnitudes in Fig. \ref{fig:fig5b}. The accuracy differences among the VPINN and PINN formulations are provided in the form of summarizing $R^{2}$ metrics in Table \ref{tab:r2_1}.

\begin{table}[h!]
	\begin{center}
		\begin{tabular}{c  c  c  c  c}
			\hline
			\multirow{3}{10em}{Test case}   &\multicolumn{4}{c}{Coefficient of determination $R^{2}$} \\ 
			&\multicolumn{2}{c}{$u_{1}$}  &\multicolumn{2}{c}{$u_{2}$} \\
                & $w_{3}=0$ & $w_{3}=1$ & $w_{3}=0$ & $w_{3}=1$ \\
			\hline
			VPINN ($n=0$)  & $0.991$ & $0.996$ & $0.97$   & $0.995$
			\\   
			PINN ($n=0$)  & $0.974$ & $0.974$ & $0.94$   & $0.95$
			\\               
			\hline
		\end{tabular}
	\end{center}
	\caption{Coefficient of determination $R^{2}$ for the test cases considering one circular inclusion ($n=0$). Comparison between the classical PINN and the proposed VPINN formulations.}
	\label{tab:r2_1}
\end{table}

\subsection{Single flower-type second-phase}
\label{sec:FlowPhase}
In the sequel, a flower-shaped second phase ($n=2$, Eq. \eqref{eq:3a}) within a plane-stressed square plate $[0,1] \times [0,1]$ is investigated. Figure \ref{fig:fig6}a depicts the set of 2356 integration cells for a 7-level quadtree, as well as the distribution of the integration points, the boundary points, and the auxiliary domain points. The location and target displacement values of the boundary points are obtained by implementing the GBEM method with a boundary mesh of 274 quadratic line elements, while a $5 \times 5$ grid of points inside $\Omega_{0}$ is used to generate the inner domain data $D_{\Omega}$. The evolution of the normalized loss function $\mathcal{L}(\boldsymbol{\lambda})/ \mathcal{L}_{0}$ is provided in Fig. \ref{fig:losses}b.
\begin{figure}[H]
\centering
\includegraphics[scale=0.55]{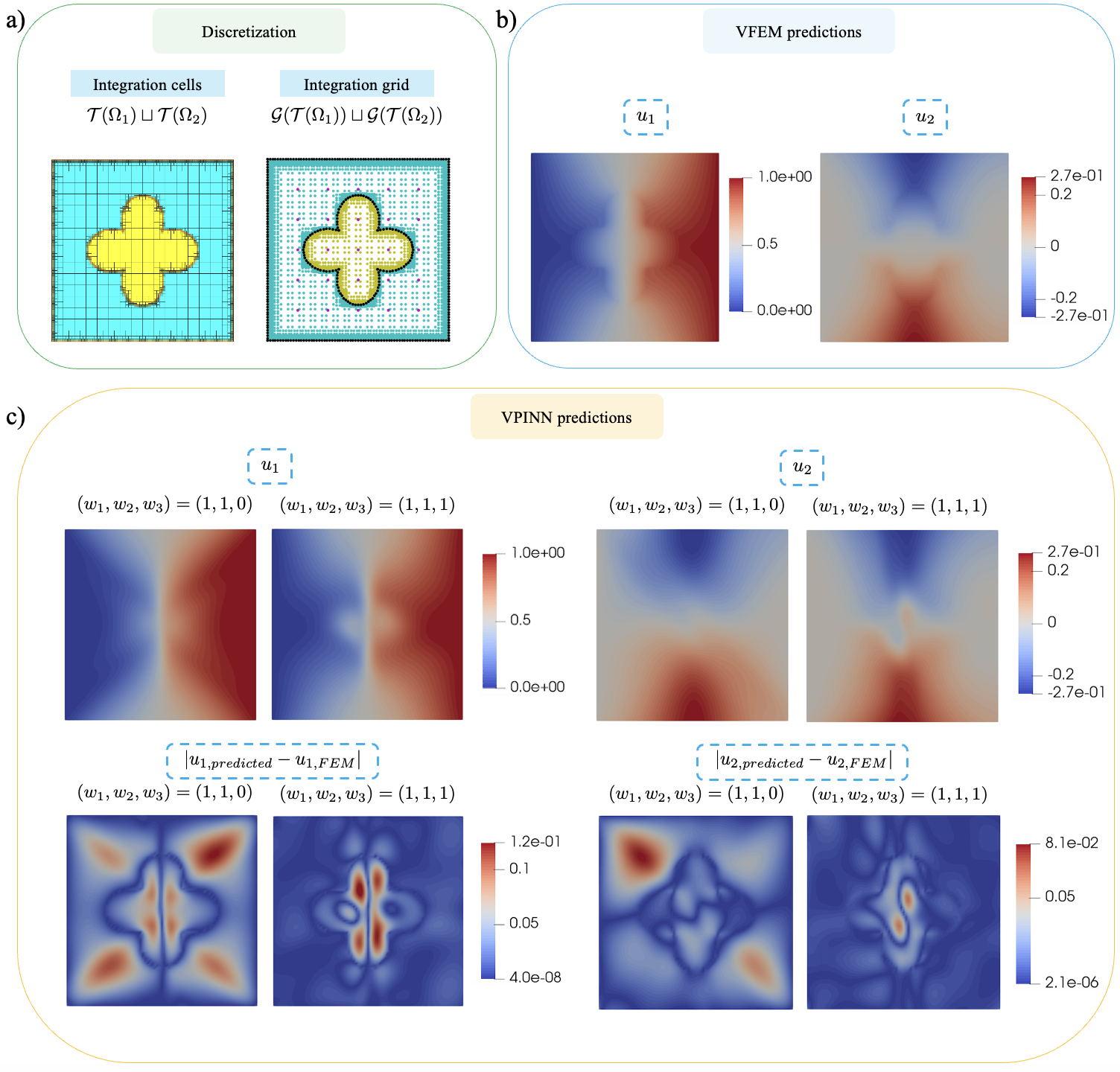} 
\caption{Plane-stress plate with one flower-shaped inclusion ($n=2$): a) Integration cells and training point grid, b) Numerical results for the displacement components acquired from the VFEM formulation, and, c) Contour plots of the displacement prediction of the VPINN formulation with ($w_{3}=1$) and without ($w_{3}=0$) taking into account target values on the additional auxiliary interior points. The contour plots of the point-wise absolute errors are provided as well using a common scale in the colormap.}
\label{fig:fig6}
\end{figure}

\begin{figure}[H]
\centering
\includegraphics[scale=0.5]{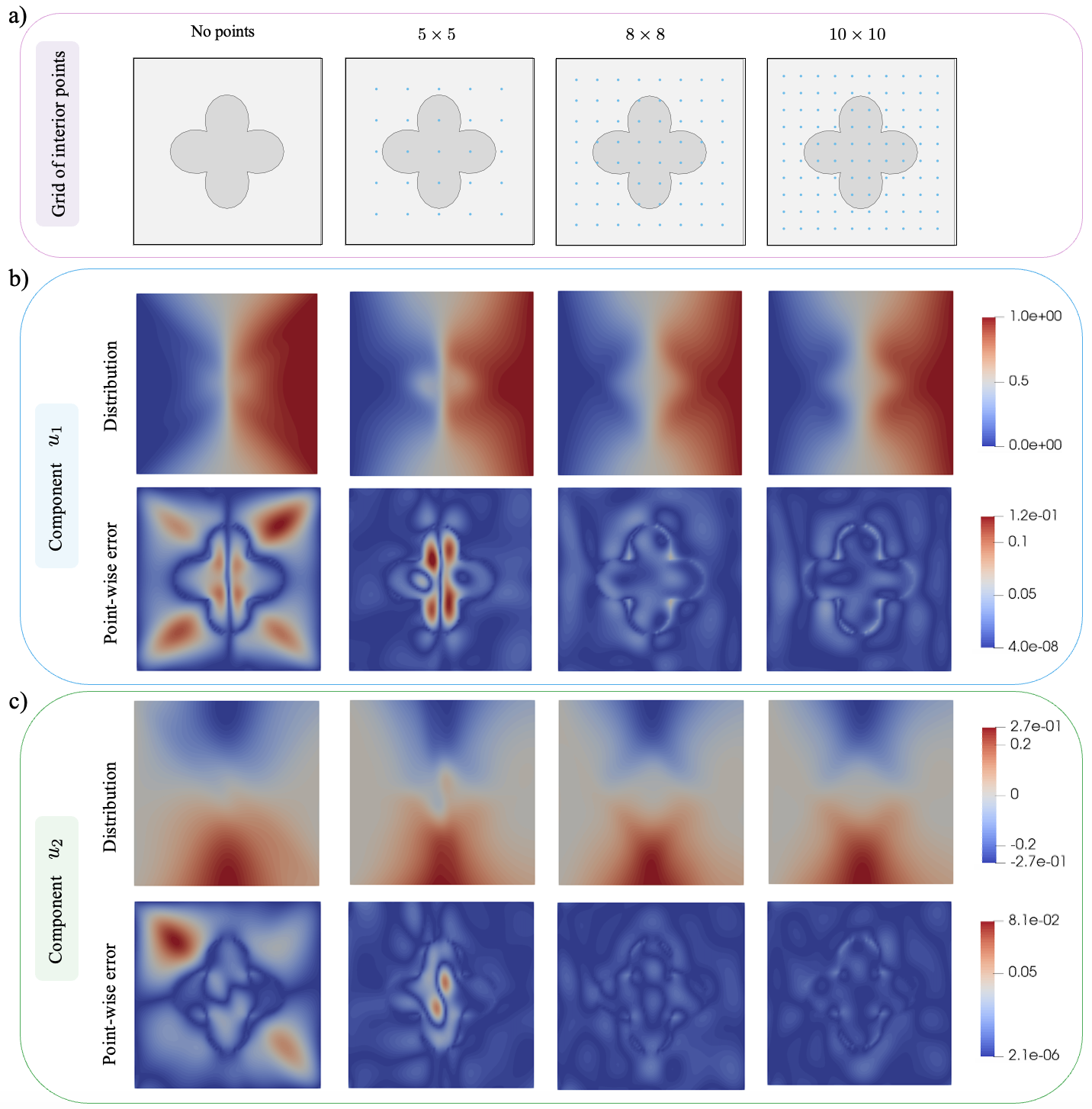} 
\caption{Sensitivity analysis for the case with the one flower-shaped inclusion ($n=2$) with respect to the effect of the auxiliary interior data points: a) the different interior point grid cases, b) contour plots of $u_{1}$ component of displacement and point-wise error, c) contour plots of $u_{2}$ component of displacement and point-wise error. The comparisons are made with respect to the numerical solution of VFEM formulation.}
\label{fig:sensitivity}
\end{figure}
The results of Fig. \ref{fig:fig6}c reveal an analogous performance with the one recorded in the circular second phase case (Section \ref{sec:Circ_SP}), with the deep VPINN predictions to capture the displacement field with a considerable accuracy, already in the case that no internal data are considered ($w_{3}=0$). However, displacement field errors within both material phases are recorded, their magnitude exceeding 10$\%$ at different internal positions. A substantial improvement in the accuracy of the computations takes place upon the introduction of auxiliary interior data point $D_{\Omega}$ information. In particular, the field approximations of both $u_{1}$ and $u_{2}$ are improved in $\Omega_{1}$, with the pointwise errors to concentrate solely in the $\Omega_{2}$ phase ($w_{3}=1$, Fig. \ref{fig:fig6}c). 

Noting that the maximum pointwise displacement errors in the second material phase may still exceed 10$\%$ for a $5 \times 5$ grip of internal points, a sensitivity analysis of the number of auxiliary interior data in $D_{\Omega}$ is performed. In particular, the VPINN model predictions with grids consisting of $5 \times 5$, $8 \times 8$, and $10 \times 10$ internal points are compared, adding information through the incorporation of approximately 5, 11, and 18$\%$ more inner datapoints, compared to the $w_{3} = 0$ case, where no internal information is considered. The relevant results indicate a significant reduction in the pointwise errors, both for the horizontal $u_{1}$ and vertical $u_{2}$ displacement components, with respect to the reference $w_{3} = 0$ case (Fig. \ref{fig:sensitivity}c). In particular, the contour plots of $u_{1}$ and $u_{2}$ (Fig. \ref{fig:sensitivity}b) suggest that an inner grid of $\#D_{\Omega} = 8 \times 8$ points leads to a reduction of the maximum magnitude of the relative pointwise error to 5$\%$. The addition of further internal datapoints ($\#D_{\Omega} = 10 \times 10$ case) which nearly doubles the intermal information provided, does not substantially improve the analysis results, with the VPINN predictions to lie in close agreement with the VFEM results over the entire domain (Fig. \ref{fig:fig6}).

\subsection{Multiple non-uniform second-phases}
\label{sec:RandomSecondPhase}

The VPINN performance is subsequently analyzed for the case of multiple disjoined inner second material phases. A pattern containing four arbitrarily positioned, non-overlapping circular second phases ($n=0$, Eq. \eqref{eq:3a}) is initially considered. A domain with double the size of the single, second material phase case is employed $[0,2] \times [0,2]$, with the $\Omega_{2}^{i}$ phase centers placed at $\mathbf{c}^{1} = (0.5,0.45)$, $\mathbf{c}^{2} = (1.3,0.65)$, $\mathbf{c}^{3} = (1.5,1.55)$ and $\mathbf{c}^{4} = (0.5,1.3)$, the coordinates computed so that they violate the classical Thomson's sphere placement problem \cite{Karathan_Bayesian}.  The employed 7-level quadtree segmentation leads to a set of 2911 integration cells with a $2\times2$ Gauss-Legendre grid. The training boundary data $D_{\partial\Omega}$ and domain data $D_{\Omega}$ are acquired by the GBEM computations, considering 332 boundary elements and a grid of $10 \times 10$ interior points (Fig. \ref{fig:fig7}a). The normalized loss function $\mathcal{L}(\boldsymbol{\lambda})/ \mathcal{L}_{0}$ is provided in Fig. \ref{fig:losses}c.

\begin{figure}[H]
\centering
\includegraphics[scale=0.55]{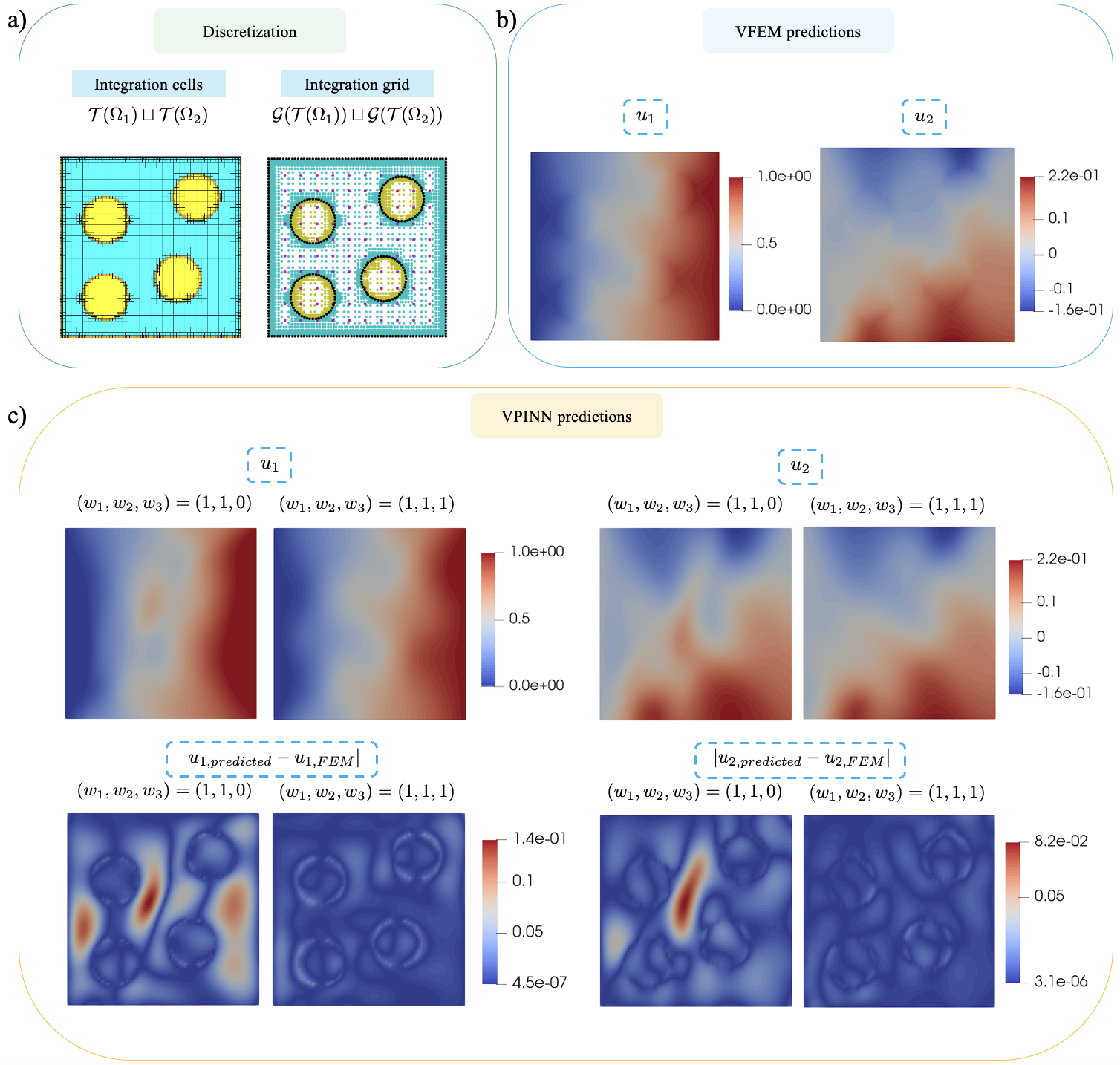} 
\caption{Plane-stress plate with 4 distributed circular inclusions ($n=0$): a) Integration cells and training point grid, b) Numerical results for the displacement components acquired from the VFEM formulation, and, c) Contour plots of the displacement prediction of the VPINN formulation with ($w_{3}=1$) and without ($w_{3}=0$) taking into account target values on the additional auxiliary interior points. The contour plots of the point-wise absolute errors are provided as well using a common scale in the colormap.}
\label{fig:fig7}
\end{figure}

The contour plots of Fig. \ref{fig:fig7}c reveal an overall satisfactory performance of the VPINN model in the absence of inner data ($w_{3}=0$, Eq. \eqref{eq:221e2}), with the displacement field errors to concentrate at the material parts of the first phase that are positioned among the circular $\Omega_{2}$ domain parts. A significant improvement on the VPINN predictions is obtained by the addition of the domain data $D_{\Omega}$ information in the loss function (Fig. \ref{fig:fig7}c). In that case,  the pointwise error is substantially reduced, with the VPINN predictions at the intermediate spaces of the second material phase approaching the VFEM predictions within a 5$\%$ accuracy threshold.

\begin{figure}[H]
\centering
\includegraphics[scale=0.55]{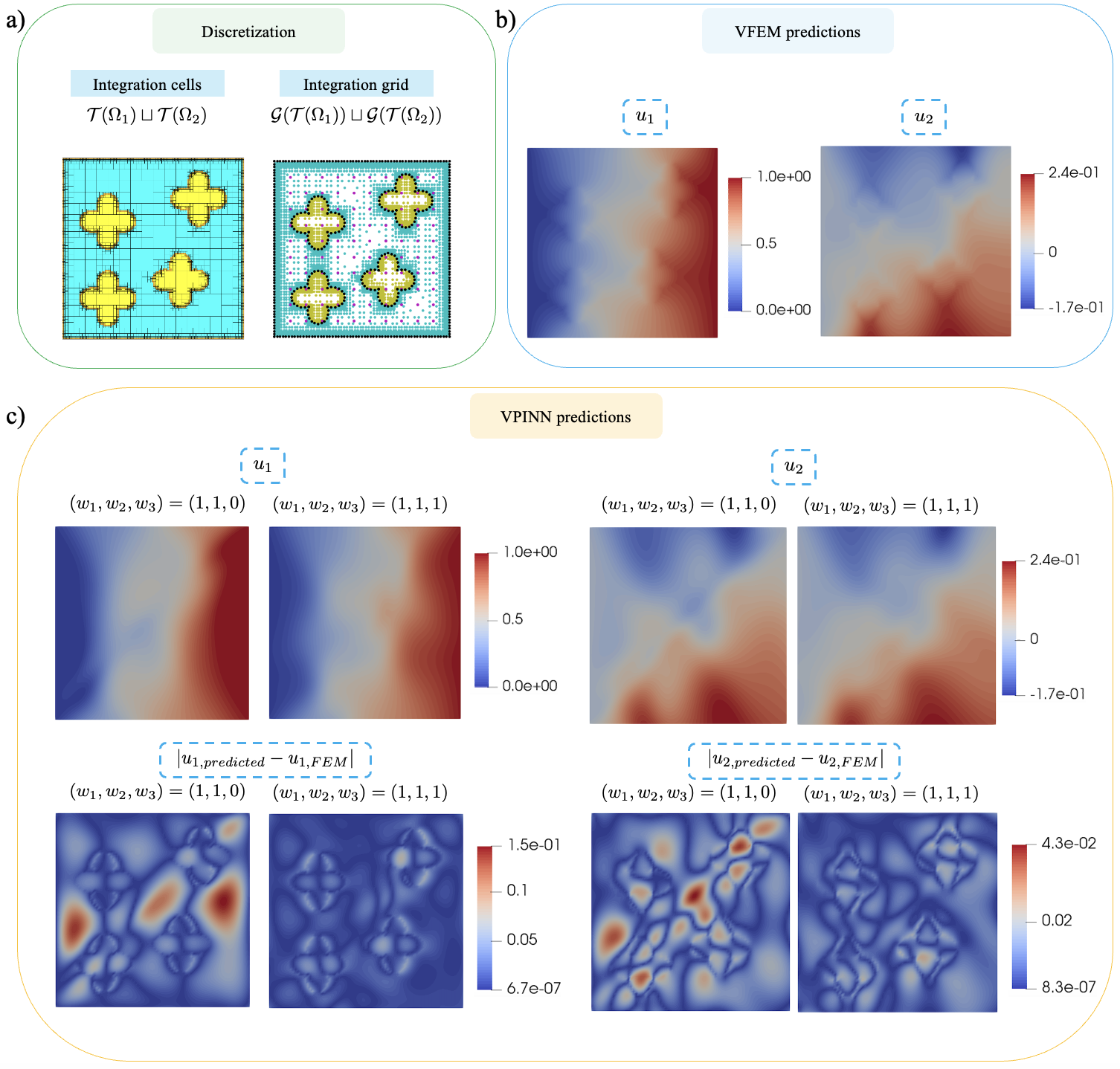} 
\caption{Plane-stress plate with 4 distributed flower-shaped inclusions ($n=2$): a) Integration cells and training point grid, b) Numerical results for the displacement components acquired from the VFEM formulation, and, c) Contour plots of the displacement prediction of the VPINN formulation with ($w_{3}=1$) and without ($w_{3}=0$) taking into account target values on the additional auxiliary interior points. The contour plots of the point-wise absolute errors are provided as well using a common scale in the colormap.}
\label{fig:fig8}
\end{figure}

Subsequently, flower-shaped second-phase material patterns, with the same distribution as the one previously employed, are analyzed, considering a set of 3283 integration cells with a $2times2$ Gauss-Legendre grid. An boundary mesh of 332 quadratic boundary elements along with a grid of $10 \times 10$ internal domain points is employed for the generation of the $D_{\partial\Omega}$ and $D_{\Omega}$ data points. In Fig. \ref{fig:losses}d the evolution of the normalized loss function $\mathcal{L}(\boldsymbol{\lambda})/ \mathcal{L}_{0}$ is provided, while the domain discretization information along with the VFEM predictions are summarized in Figure \ref{fig:fig8}a and Figure \ref{fig:fig8}b, respectively. Comparative metrics of the performance of the VPINN formulation for all second-phase patterns are provided in Table \ref{tab:r2_2}.

\begin{table}[h!]
	\begin{center}
		\begin{tabular}{c  c  c  c  c}
			\hline
			\multirow{3}{10em}{Test case}   &\multicolumn{4}{c}{Coefficient of determination $R^{2}$} \\ 
			&\multicolumn{2}{c}{$u_{1}$}  &\multicolumn{2}{c}{$u_{2}$} \\
                & $w_{3}=0$ & $w_{3}=1$ & $w_{3}=0$ & $w_{3}=1$ \\
			\hline
                ($n=0$)  & $0.98$ & $0.99$ & $0.95$   & $0.973$
                \\
			($n=2$)  & $0.98$ & $0.995$ & $0.95$   & $0.992$
			\\   
			$4 \times (n=0)$  & $0.98$ & $0.99$ & $0.98$   & $0.99$
			\\               
                $4 \times (n=2)$  & $0.983$ & $0.998$ & $0.96$   & $0.997 $
                \\            
			\hline
		\end{tabular}
	\end{center}
	\caption{Coefficient of determination $R^{2}$ for the test cases considering one circular inclusion ($n=0$). Comparison between the classical PINN and the proposed VPINN formulations.}
	\label{tab:r2_2}
\end{table}

The deep learning predictions of Fig. \ref{fig:fig8}c indicate an overall satisfactory representation of the displacement field in the absence of internal point information ($w_{3}=0$, Fig. \ref{fig:fig8}b). As in the circular second-phase case (Fig. \ref{fig:fig9}), low performance is recorded in the areas between the second-phase, flower-shaped, designs. These pointwise deviations are significantly reduced when the information of the additional training data $D_{\Omega}$ is taken into consideration. In that case, the maximum of the relative pointwise prediction errors is concentrated at the domain interfaces, where sharp displacement transitions are recorded, the relative differences lying below 3$\%$ for the $u_2$ displacement component. 

The VPINN model is capable of capturing the horizontal $u_1$ displacement component, with a coefficient of determination that exceeds 98$\%$ for all cases, without the use of internal point information (Table \ref{tab:r2_2}). The corresponding vertical displacement prediction capability is overall lower, lying in the order of 95-96 $\%$ for all patterns, except for the case of non-uniform, circular second phases, where higher values are reported (Table \ref{tab:r2_2}). The addition of inner domain information enhances the prediction accuracy for both displacement components, the effect being more prominent in the initially, worse-performing, $u_2$ displacement component (Table \ref{tab:r2_2}).

\begin{figure}[H]
\centering
\includegraphics[scale=0.55]{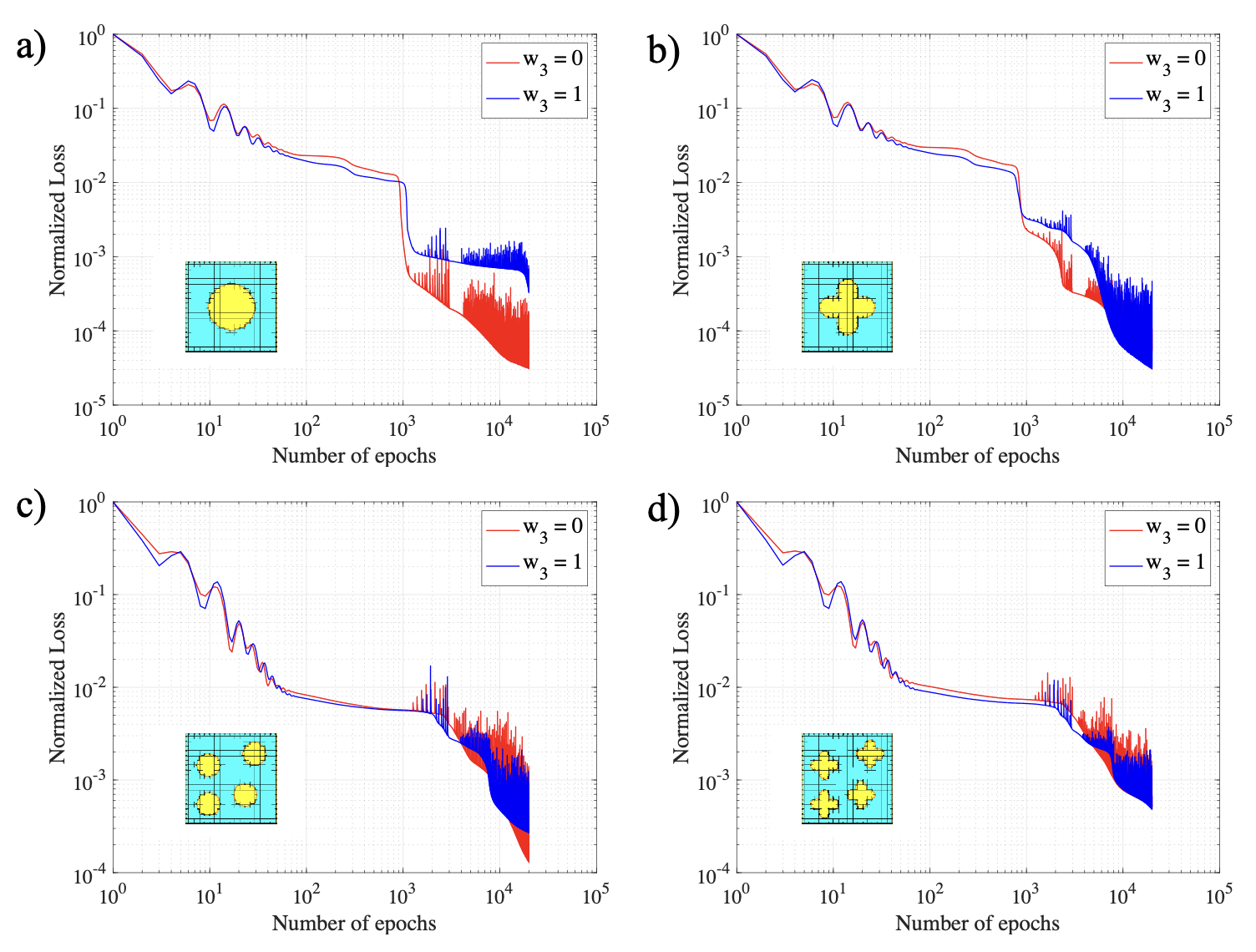} 
\caption{Loss function as a function of the number of epochs for the training procedures: with ($w_{3} = 0$) and without ($w_{3} = 1$) the use of interior point, a) circular inclusion of Fig. \ref{fig:fig5}, flower-shaped inclusion of Fig. \ref{fig:fig6}, pattern of 4 circular inclusions of Fig. \ref{fig:fig7} }
\label{fig:losses}
\end{figure}

\subsection{Tree-based integration and computational efficiency}
\hspace{\parindent}
The computational efficiency of the tree-based VPINN formulation is analyzed considering the same training data sets $D_{\partial \Omega}$ and data points $D_{\Omega}$ employed in Section \ref{sec:RandomSecondPhase}, along with a dense integration grid with equal element sizes over both material directions (Fig. \ref{fig:fig9}b). The integration cells of the dense grid (Fig. \ref{fig:fig9}b) are equal in size to the smallest integration cell obtained from the 7-level quadtree structure (Fig. \ref{fig:fig9}a) for comparison purposes. For the dense grid of 5833 cells, a $2\times2$ Gauss-Legendre rule is implemented in each one.

\begin{figure}[H]
\centering
\includegraphics[scale=0.5]{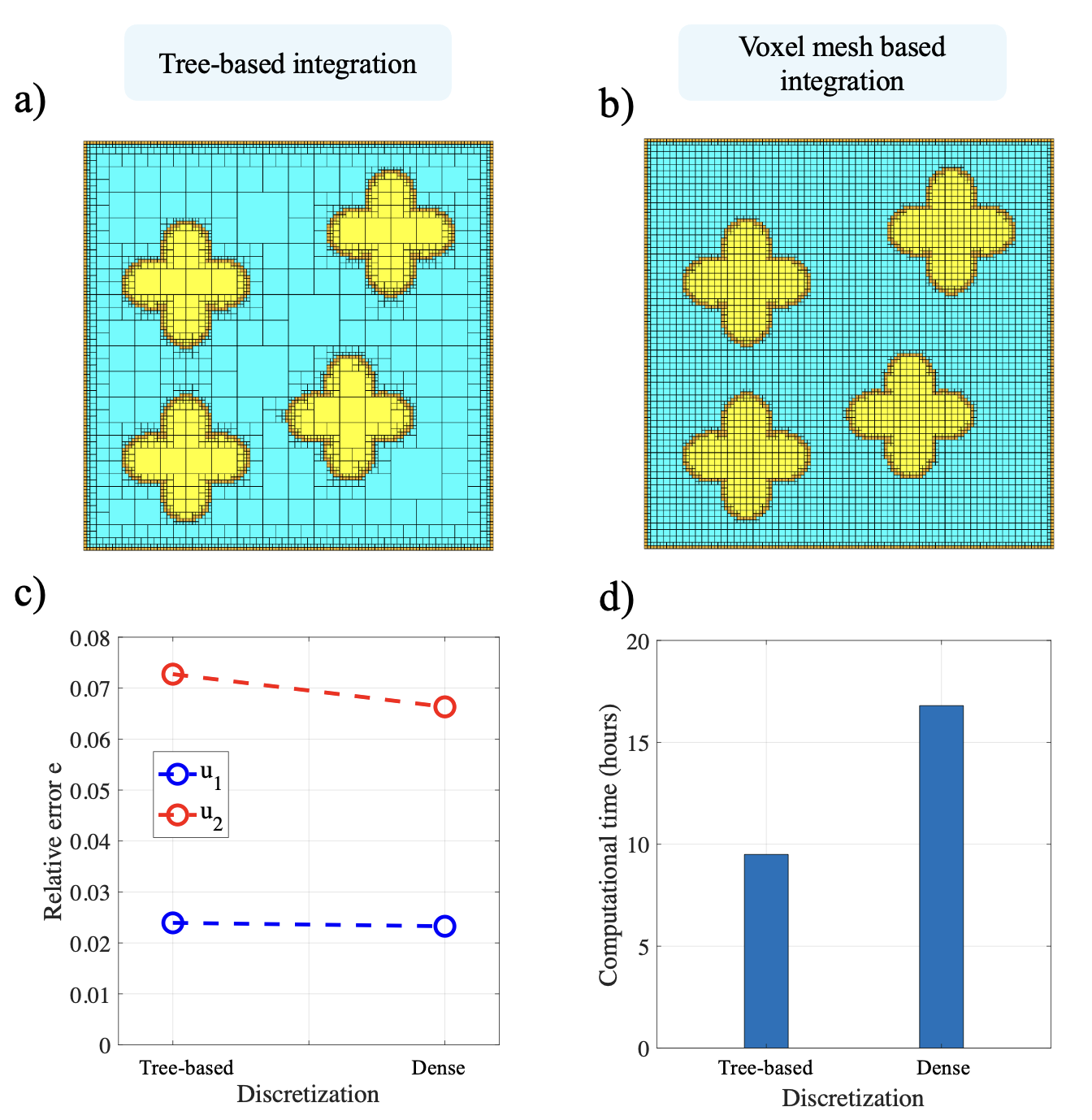} 
\caption{Integration cells of the numerical test case 4, considering a) a quadtree structure of 7 levels, b) a dense discretization with a cell size equal to the smallest cell (leaf) of the quadtree. c) The relative error (Eq. x) for the displacement components $u_{1}$ and $u_{2}$ for the two different discretizations and, d) the computation execution time (in hours).}
\label{fig:fig9}
\end{figure}

The relative errors $e(u_{1}; u_{FEM,1})$ and $e(u_{2}; u_{FEM,2})$ are computed using as a reference the VFEM results of Fig. \ref{fig:fig8}. Their magnitude indicates that the horizontal displacement component is practically unaffected by the substantially higher number of elements employed in the dense modeling case, while the vertical $u_2$ component accuracy is improved solely by 0.005 (Fig. \ref{fig:fig9}c), so that both predictions can be considered practically equivalent. However, as summarized in the bar plots of Fig. \ref{fig:fig9}d, the computational time of the training process is $9.5$ hours for 20000 epochs for the tree-based integration case, being approximately half than the training time of the dense discretization.

The recorded computational cost difference, while substantial, remains a fraction of the three-dimensional case, with volume integrations requiring, as a minimum, two times the integration points of the 2D space, substantially increasing the computational cost at the element scale. The observation highlights the importance of adaptive integration in the computation of the mechanics of multiphase architected materials, with the elaborated VPINN formulation to allow for sufficient numerical accuracy upon significant computational efficiency gains.

}

\section{Conclusions}
\label{Sec:Conclusions}
{ 
\hspace{\parindent}

Overall, the work has elaborated a Variational-PINN machine learning framework for the modeling of the mechanical performance of multiphase materials. The VPINN method exploits the flexibility of the Petrov-Galerkin formulation, integrating deep neural networks as trial functions, with the robust dataset generation of the GBEM for the exterior, interface boundaries, and auxiliary interior points. The framework is complemented by an adaptive tree-based integration for the evaluation of the weak form integrals. It has been found that:

\begin{enumerate}
\item The VPINN ML framework is capable of capturing the deformation field of multiphase materials with remarkable accuracy, with the performance improved upon the integration of additional auxiliary point information. 

\item For the same NN architecture and training settings (optimizer settings, number of training epochs), the elaborated VPINN yields an overall better performance in the modeling of the mechanics of multiphase materials, compared to classical, strong PINN formulations.

\item The applied tree-based integration for the weak form integrals offers the advantage of reducing the training computational cost, without sacrificing the accuracy of the numerical predictions. The formulation is particularly important for multiphase material architectures, with complex inner architectures, and discrete displacement fields for each material phase. Moreover, its reduced computational cost offers the possibility to incorporate \textit{p} refinement strategies and polynomial orders as test functions of each integration cell.

\item The combination of the VPINN formulation with the Boundary Element Method provides the advantage of the direct generation of outer boundary, interface, and inner domain data, without the need for volume meshing.

\end{enumerate}

The findings of the present study aspire to provide a reference for the development of VPINNs for the mechanical analysis of multiphase materials. The formulation can be readily extended to the three-dimensional cases, where computational cost reductions are expected to be more pronounced. The use of different types and orders of polynomial test functions, as well as different NNs per material constituent are aspects worthy of further exploration.

}

\section*{Acknowledgements}
{ 
The authors would like to gratefully acknowledge the support of the computational resources of the High Performance Computing Center of the New York University in the Abu Dhabi campus.
}

\section*{Appendix A} \label{appA}
{
\textbf{\large Sensitivity analysis with respect to the neural network architecture}
\vspace{1cm}

Considering the Test case 1 with the one circular inclusion $n=0$, a sensitivity analysis with respect to the hyperparameters that control the network architecture, the number of hidden layers $N_{l}$ and the number of neurons per layer $N_{n}$, is conducted for the VPINN and classical PINN formulations. The accuracy of the numerical predictions on the testing set $T_{\Omega}$ is assessed calculating the coefficient of determination $R^{2}$ (Eq. \eqref{eq:232e1}) for each displacement component $u_{i}$ as well as the magnitude $|\mathbf{u}|$.

To begin with, the sensitivity of the formulations is tested on different number of hidden layers $N_{l}$, fixing the number of neurons per layer to $N_{n} = 60$. The calculated values of $R^{2}$ for $u_{i}$ and $|\mathbf{u}|$ are provided in the Table \ref{tab:sens_1}, revealing the better performance of the VPINN compared to PINN, for every considered $N_{l}$. The best accurary of the VPINN formulation is recorded for $N_{l} = 5$, which defines the architecture that is employed in the numerical test cases of Section 3.
\begin{table}[h!]
	\begin{center}
		\begin{tabular}{c  c  c  c  c  c  c  c}
			\hline
			\multirow{3}{10em}{Number of hidden layers $N_{l}$} &\multirow{3}{10em}{Number of trainable parameters}   &\multicolumn{4}{c}{Coefficient of determination $R^{2}$}
                \\ 
			& &\multicolumn{3}{c}{VPINN}  &\multicolumn{3}{c}{PINN} 
                \\
                & &\multicolumn{1}{c}{$u_{1}$}  &\multicolumn{1}{c}{$u_{2}$}  &\multicolumn{1}{c}{$|\mathbf{u}|$} &\multicolumn{1}{c}{$u_{1}$}   &\multicolumn{1}{c}{$u_{2}$} &\multicolumn{1}{c}{$|\mathbf{u}|$}
                \\
			\hline
			3  & 7622  & $0.986$ & $0.978$ & $0.989$  & $0.974$ & $0.941$ & $0.975$
			\\   
			4  & 11282 & $0.987$ & $0.946$ & $0.988$  & $0.974$ & $0.972$ & $0.975$
			\\               
                5  & 14942 & $0.991$ & $0.968$ & $0.993$  & $0.974$ & $0.941$ & $0.975$
                \\   
                8  & 25922 & $0.988$ & $0.964$  & $0.989$ & $0.972$ & $0.939$  & $0.973$
                \\                   
			\hline
		\end{tabular}
	\end{center}
	\caption{Coefficient of determination $R^{2}$ for the test cases considering one circular inclusion ($n=0$). Comparison between different number of hidden layers $N_{l}$ prescribing the number of neurons per layer $N_{n} = 60$.}
	\label{tab:sens_1}
\end{table}
In the sequel, the sensitivity of the formulations is assessed on different number of neurons per layer $N_{n}$, fixing the number of hidden layers to $N_{l} = 5$. Table \ref{tab:sens_2} provides the calculated values of $R^{2}$ for $u_{i}$ and $|\mathbf{u}|$, showing again a superior performance of the VPINN compared to PINN, for every considered $N_{n}$. This test also reveals that the architecture $N_{l} = 5$, $N_{n} = 60$ leads the best accurary of the VPINN formulation.
\begin{table}[h!]
	\begin{center}
		\begin{tabular}{c  c  c  c  c  c  c  c}
			\hline
			\multirow{3}{10em}{Number neurons per layer $N_{n}$} &\multirow{3}{10em}{Number of trainable parameters}   &\multicolumn{4}{c}{Coefficient of determination $R^{2}$}
                \\ 
			& &\multicolumn{3}{c}{VPINN}  &\multicolumn{3}{c}{PINN} 
                \\
                & &\multicolumn{1}{c}{$u_{1}$}  &\multicolumn{1}{c}{$u_{2}$}  &\multicolumn{1}{c}{$|\mathbf{u}|$} &\multicolumn{1}{c}{$u_{1}$}   &\multicolumn{1}{c}{$u_{2}$} &\multicolumn{1}{c}{$|\mathbf{u}|$}
                \\
			\hline
			20  & 1782  & $0.987$ & $0.927$ & $0.99$  & $0.974$ & $0.939$ & $0.975$
			\\   
			40  & 6762 & $0.989$ & $0.972$ & $0.99$  & $0.975$ & $0.947$ & $0.975$
			\\               
                60  & 14942 & $0.991$ & $0.968$ & $0.993$  & $0.974$ & $0.941$ & $0.975$
                \\   
                80  & 26322 & $0.990$ & $0.968$  & $0.992$ & $0.974$ & $0.945$  & $0.975$
                \\                   
			\hline
		\end{tabular}
	\end{center}
	\caption{Coefficient of determination $R^{2}$ for the test cases considering one circular inclusion ($n=0$). Comparison between different number of neurons per layer $N_{n}$ prescribing the number of hidden layers $N_{l} = 5$.}
	\label{tab:sens_2}
\end{table}

The last step of the sensitivity analysis is to assess the effect of the "the shape" of the NN in terms of "length" ($N_{l}$) and "thickness" ($N_{n}$), fixing the number of total trainable parameters. Table \ref{tab:sens_3} provides the $R^{2}$ for 4 different network architectures: $N_{l} \times N_{n} = 3 \times 80$, $N_{l} \times N_{n} = 5 \times 60$, $N_{l} \times N_{n} = 10 \times 40$ and $N_{l} \times N_{n} = 16 \times 30$. All these architectures have the comparable number of trainable parameters, $\cong 14900$. Regarding the $u_{1}$ component, the lowest value of $R^{2}$ is recorded for the network $N_{l} \times N_{n} = 16 \times 30$, which is has the highest "length" to "thickness" ratio ("$N_{l}/N_{n}$"). However, regarding $u_{2}$, the $R^{2}$ takes the minimum value for $N_{l} \times N_{n} = 3 \times 80$, which has the lowest "length" to "thickness" ratio. Moreover, the $R^{2}$ value for the displacement magnitude $|\mathbf{u}|$ takes the highest value for $N_{l} \times N_{n} = 5 \times 60$, which is the architecture used in the numerical examples of Section 3.
\begin{table}[h!]
	\begin{center}
		\begin{tabular}{c  c  c  c}
			\hline
			\multirow{2}{10em}{Network set up ($N_{l} \times N_{n}$)}   &\multicolumn{3}{c}{Coefficient of determination $R^{2}$} \\ 
			&\multicolumn{1}{c}{$u_{1}$}  &\multicolumn{1}{c}{$u_{2}$}  &\multicolumn{1}{c}{$|\mathbf{u}|$} \\
			\hline
			$3 \times 80$  & $0.988$ & $0.931$  & $0.990$
			\\   
			$5 \times 60$  & $0.991$ & $0.968$  & $0.993$
			\\               
                $10 \times 40$  & $0.986$ & $0.973$  & $0.987$
                \\            
                $16 \times 30$  & $0.984$ & $0.971$  & $0.985$
                \\
			\hline
		\end{tabular}
	\end{center}
	\caption{Coefficient of determination $R^{2}$ for the test cases considering one circular inclusion ($n=0$). Comparison between different number of network architectures $N_{l} \times N_{n}$ prescribing almost the same number of trainable parameters.}
	\label{tab:sens_3}
\end{table}

}
 \clearpage

\bibliographystyle{unsrt}

\bibliography{References}

\end{document}